\documentclass[final,5p,times,twocolumn,authoryear]{elsarticle}

\usepackage{array}
\usepackage{amsmath,epsfig,graphicx,amsfonts}
\usepackage[hang,raggedright,nooneline]{subfigure}
\usepackage[caption=false,font=footnotesize]{subfig}
\usepackage{subfigmat}
\usepackage{color}

\usepackage{amssymb}
\usepackage{amsthm}

\journal{Vision Research}

\begin{document}

\begin{frontmatter}



\title{Predicting the Blur Visual Discomfort for Natural Scenes\\by the Loss of Positional Information}


\author[DIET]{Elio D. Di Claudio}
\author[DIET]{Paolo Giannitrapani}
\author[FDIET]{Giovanni Jacovitti}

\address[DIET]{Dept. of Information Engineering, Electronics and Telecommunications (DIET), University of Rome ``La Sapienza,'' Via Eudossiana 18, I-00184 Rome, Italy.}
\address[FDIET]{Formerly with DIET.}

\begin{abstract}
The perception of the blur due to accommodation failures, insufficient optical correction or imperfect image reproduction is a common source of visual discomfort, usually attributed to an anomalous and annoying distribution of the image spectrum in the spatial frequency domain.
In the present paper, this discomfort is attributed to a loss of the localization accuracy of the observed patterns. It is assumed, as a starting perceptual principle, that the visual system is optimally adapted to pattern localization in a natural environment.
Thus, since the best possible accuracy of the image patterns localization is indicated by the positional Fisher Information, it is argued that the blur discomfort is strictly related to a loss of this information.
Following this concept, a receptive field functional model, tuned to common features of natural scenes, is adopted to predict the visual discomfort. It is a complex-valued operator, orientation-selective both in the space domain and in the spatial frequency domain.
Starting from the case of Gaussian blur, the analysis is extended to a generic type of blur by applying a positional Fisher Information equivalence criterion. Out-of-focus blur and astigmatic blur are presented as significant examples. 
The validity of the proposed model is verified by comparing its predictions with subjective ratings. The model fits linearly with the experiments reported in independent databases, based on different protocols and settings.

\end{abstract}


%Research highlights
%\item \emph{To the best knowledge of authors, existing quantitative studies about the blur perception phenomenon are mostly focused on the response of the visual system to specific patterns localized in space or in frequency domains, that does not reflect the everyday visual experience.}
%
%\item \emph{Therefore, the work aims to provide a model for the realistic estimation of  the visual discomfort caused by the optical blur of natural scenes, considering  the cognitive needs of the subjects  and the adaption capability of the visual system.}
%
%\item \emph{This seems a timely issue, considering also that vision impairment caused by refractive error occupies a relevant place in the Vision 2020 WHO global initiative (https://www.who.int/publications/i/item/world-report-on-vision).}

\begin{keyword}
Visual perception \sep Blur discomfort \sep Optical correction \sep Image quality assessment.



\end{keyword}

\end{frontmatter}


\section{Introduction}
\label{sec:Introduction}

Among the various sources of non-clinical visual discomfort \citep{CONLON99,LAMBOOIJ09}, the blur caused by refractive errors is perhaps the most common one.

The discomfort associated with blur is often explained as a consequence of the concentration of the spatial energy spectrum of the perceived image into some bands, or as a byproduct of the discrepancy of this spectrum from the expected spectrum of natural images \citep{OHARE11,WILKINS16}. Alternative explanations addressed the mismatch of the spatial patterns with the expected ones \citep{KAYARGADDE96,WANG04B,BARONCINI09A}. An in-depth account of previous studies and mathematical models about the blur phenomenon is provided in \citep{WATSON11}.

Looking at a possible physical source of the blur discomfort, three hypotheses are examined in \citep{OHARE13}. The first hypothesis is that discomfort is stimulated by the weak response of the accommodation system. A second hypothesis, somewhat related to the first one, is that discomfort arises because the “micro-fluctuations” observed in the accommodation feedback signal become ineffective \citep{CHARMAN15,METLAPALLY16,MARIN-FRANCH17,CHOLEWIAK18}. A third hypothesis maintains that, when an image is correctly projected onto the retina, the receptive fields produce a parsimonious, \emph{sparse} representation of this image \citep{MCILHAGGA12}. The spatial spread caused by blur excites more receptors, producing a metabolic overload \citep{JUREVIC10}.

In the present approach, the blur is viewed as a cause of a cognitive loss, and the discomfort as the immediate consequence of this loss. It is argued that, among the basic cognitive functions of the human visual system (HVS), detection, recognition, and coarse localization functions are strongly conditioned by the individual experience. Conversely, it seems plausible that the fine localization function is committed to stabler and inter-subjective functions of the HVS.

Based on the above consideration, the present approach starts from postulating that, under normal conditions, the HVS performs the fine localization of the observed objects with the best accuracy allowed by its physical macro-structure, given the characteristics of the environment and man’s interaction.

This assumption is fundamental, because it is known from the estimation theory that the maximum accuracy attainable when measuring the fine position of patterns in background noise is deduced from the Fisher Information about positional parameters. In fact, the Fisher Information inverse yields the minimum estimation variance \citep{TREES92}.

Therefore, the focus here is on how the discomfort of blurring depends on the unwanted loss of Positional Fisher Information (PFI) on observed patterns.

The present cognitive approach is agnostic as to whether the discomfort is related to accommodation frustration or metabolic unbalances. On the other hand, it is compliant with the fact that blur discomfort concerns the regions of visual interest \citep{TAYLOR15} and that blur is not always undesired or detrimental \citep{SPRAGUE16}. For instance, blur is sometimes a wanted effect in photography and microscopy.

Previous analyses of the blur perception phenomenon were mainly oriented to the study of the visual acuity, employing specific stimuli localized either in space, such as edges, lines, crosses, or in the spatial frequency domain, such as sinusoidal gratings, or even in both domains, such as Gabor wavelets \citep{WATSON05}.

The model presented here is oriented to the evaluation of the visual impact of blur in the vision of natural scenes. To this purpose, a generic image projected on the retina is viewed as an element of the \emph{random set of natural images}, characterized by stable statistical features.

The proposed approach is based on an \emph{abstract}, \emph{functional} model of the receptive fields (RF) of the HVS, allowing for a direct computation of the PFI.

For analytical convenience, the blur is modelled as a Gaussian shaped isotropic blur. Then, the analysis is generalized to other types of blur, invoking a criterion of informational equivalence with respect to isotropic Gaussian blur under the PFI paradigm.

To verify the limits of the present approach, the model-based discomfort predictions were first compared to empirical data about the \emph{subjective} quality loss of blurred images, which is argued to be strictly related to the blur visual discomfort. These data are available in organized databases containing the results of experimental sessions finalized to Image Quality Assessment (IQA), conducted for multimedia industry purposes \citep{WANG04,BOSSE18,ITU08}. Subsequently, the model is applied to blurred images annotated with ratings of visual discomfort. The results of these experiments confirm the validity of the approach.

The paper is organized as follows. Sect.2 describes the features of the functional RF model in the space and in the spatial frequency domains. Sect.3 provides the definition of the PFI. In Sect.4, the expected Fisher Information acquired during the visual exploration of natural images is computed, and the informational equivalence of a generic blur with a Gaussian isotropic blur is stated. In Sect.5, the measure of the visual discomfort is defined. Sect.6 presents the comparison of this measure with subjective ratings. Some remarks are provided in Sect.7. Conclusions are finally drawn in Sect.8.

\section{The virtual receptive field model}
\label{sec:The virtual receptive field model}

\begin{figure}[!htb]
\centering
\includegraphics[width=3.1in]{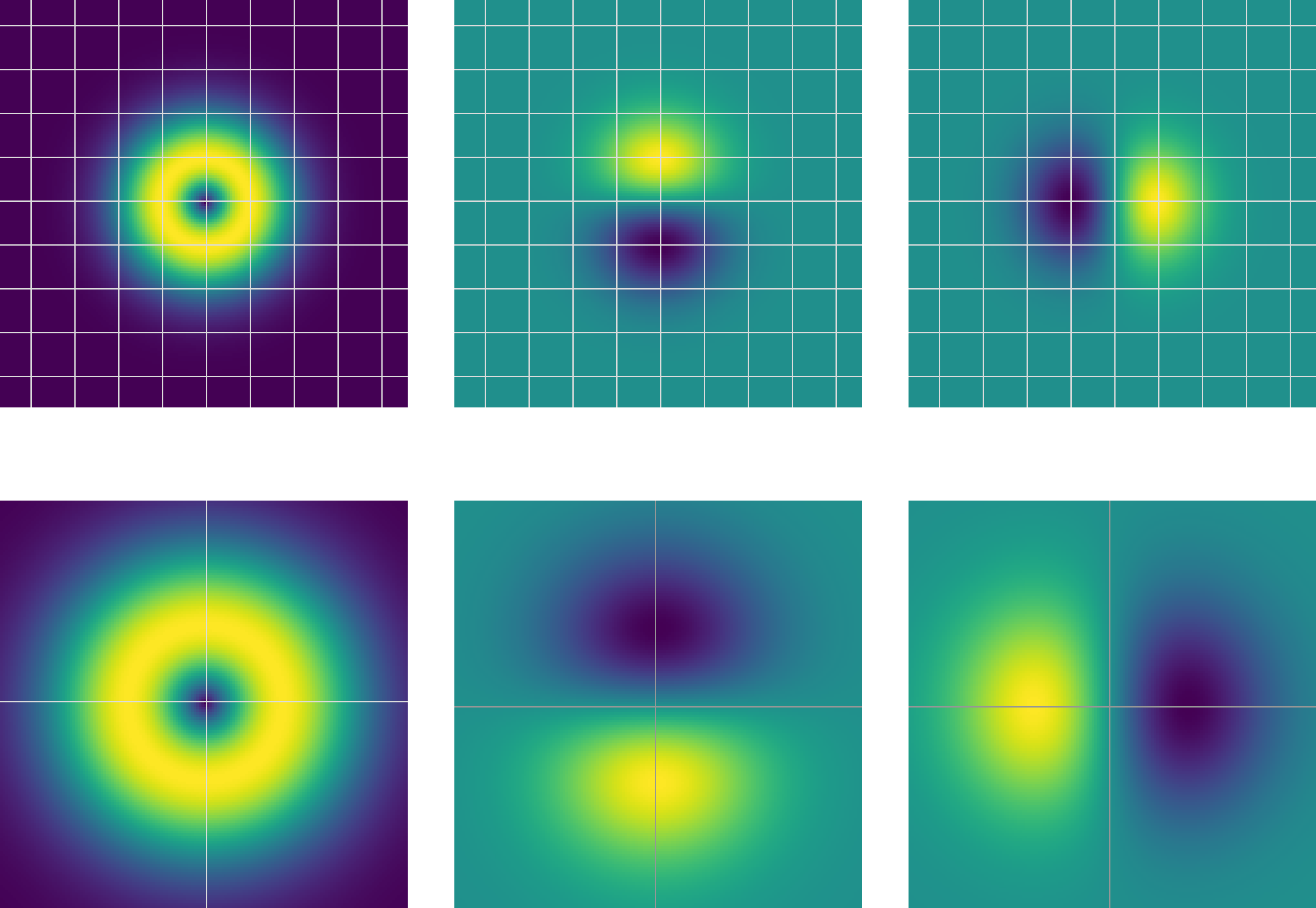}
\caption{Upper row: the magnitude of the VRF, its real and imaginary parts referred to the ideal retina grid. Lower row: the magnitude of the VTNF, its real and imaginary parts in the spatial frequency plane. The vertical and horizontal spatial frequencies span the $\left(-30,30\right) cycles/degree$ interval.}
\label{fig1}
\end{figure}

As the luminance plays a dominant role for the localizability of patterns, for the sake of simplicity only the luminance component of the images is accounted for.

The retina in the \emph{foveal vision} is abstractly modeled as a rectangular grid of receptors, whose position  is individuated by the coordinate pair $\mathbf{p}\equiv(x_1,x_2)$. Receptors are regularly spaced \emph{one arcmin} apart. The density of 60 $receptors/degree$ assures that all the image information within the 30 (=60/2) $cycles/degree$ spatial bandwidth is captured by the retina, according to the Nyquist sampling rule.

For any $\mathbf{p}$, the RF calculates a weighted sum of the luminance $\mathbf{I\left(\mathbf{p}\right)}$ in a neighborhood of $\mathbf{p}$, yielding a \emph{visual map} $y\left(\mathbf{p}\right)$. This operation corresponds mathematically to a spatial \emph{convolution}, indicated by the symbol $*$, between $I\left(\mathbf{p}\right)$ and the visual map of a single lighting point in the dark, indicated by $h(\mathbf{p})$ and referred to as the Point Spread Function (PSF) of the RF:
\begin{equation}
y\left(\mathbf{p}\right)=I\left(\mathbf{p}\right)\ast h(\mathbf{p})\; .
\label{eqn:CGG}
\end{equation}

The RF model considered here is a harmonic angular filter (HAF). HAF filters are complex valued functions, i.e., they represent \emph{pairs of real filters} \citep{JACOVITTI90}.

With reference to the spatial frequency domain, defined by the horizontal and vertical frequencies $f_1$, $f_2$, and to the polar coordinates $\rho=\sqrt{f_1^2+f_2^2}$ and $\vartheta=\arctan\left(\displaystyle\frac{f_2}{f_1}\right)$, where $\rho$ \citep{DAUGMAN80} is referred to as the \emph{radial frequency} and $\vartheta$ is the azimuth, the Fourier spectrum $H(\rho,\vartheta)$ of the RF model is defined as
\begin{equation}
H(\rho,\vartheta)=j2\pi\left(\rho e^{-S_G^2\rho^2}\cdot e^{j\vartheta}\right)
\label{eqn:VNTF}
\end{equation}
where $j$ is the imaginary unit, and $s_G$ is a parameter. This spectrum is \emph{polar separable}, i.e., it is the product of a function of the radial frequency, and a function of the azimuth. One outstanding feature of this RF model is that it is also polar separable in the space domain \citep{DAUGMAN83}. In fact, taking the inverse Fourier transform of $H(\rho,\vartheta)$, with reference to the space polar coordinates $r=\sqrt{x_1^2+x_2^2}$ and $\varphi=tg^{-1}\displaystyle\frac{x_2}{x_1}$, the PSF of the RF is:
\begin{equation}
h(r,\varphi)=-2\frac{\pi^3}{s_G^4}\cdot\left(re^{-\frac{r^2\pi}{{s_G}^2}}\cdot e^{j\varphi}\right)
\label{eqn:PSF}
\end{equation}
which has the same shape of the VNTF, except for a scale factor.

This RF is used herein as a \emph{functional} spatial vision model and, for this reason, it is referred to as \emph{Virtual Receptive Field} (VRF). For short, the term “VRF” will be used in the following to indicate also its PSF and its HAF shape, whereas its Fourier transform $H(\rho,\vartheta)$ will be referred to as Virtual Neural Transfer Function (VNTF), because it represents the \emph{spatial frequency response} of the VRF.

The magnitude and the real components of the VRF are displayed in the upper row of Fig.\ref{fig1}, where the ideal retinal grid is shown in the background. In the same figure, the magnitude, the real and the imaginary parts of the VNTF are displayed in the lower row.

%\begin{figure}[!htb]
%\centering
%\includegraphics[width=1.1in]{HGL_VRF_00.png}
%\includegraphics[width=1.1in]{HGL_01.png}
%\includegraphics[width=1.1in]{HGL_10.png}\vspace{3mm}
%\includegraphics[width=1.1in]{HGL_VRF_00_freq.png}
%\includegraphics[width=1.1in]{HGL_01_freq.png}
%\includegraphics[width=1.1in]{HGL_10_freq.png}
%\caption{Upper row. (left) $\left|h(x_1,x_2)\right|$, (center) $Re{\left\{h(x_1,x_2)\right\}}$, (right) $Im{\left\{h(x_1,x_2)\right\}}$. They represent, respectively, the magnitude of the PSF of the VRF, its real and imaginary parts, referred to the $1$ arcmin ideal retina grid.
%Lower row. (left) $\left|H(f_1,f_2)\right|$, (center) $Re{\left\{H(f_1,f_2)\right\}}$, (right) $ Im{\left\{H(f_1,f_2)\right\}}$. They represent, respectively, the magnitude of the VTNF, its real and imaginary parts in the spatial frequency plane. The vertical and horizontal spatial frequencies span the $\left(-30,30\right)$ cycles/degree range.}
%\label{fig1}
%\end{figure}

%\begin{figure}[!htb]
%\centering
%\includegraphics[width=3.1in]{HGLall.png}
%\caption{Upper row: the magnitude of the VRF, its real and imaginary parts referred to the ideal retina grid. Lower row: the magnitude of the VTNF, its real and imaginary parts in the spatial frequency plane. The vertical and horizontal spatial frequencies span the $\left(-30,30\right) cycles/degree$ interval.}
%\label{fig1}
%\end{figure}

\begin{figure}[!htb]
\centering
\includegraphics[width=3.1in]{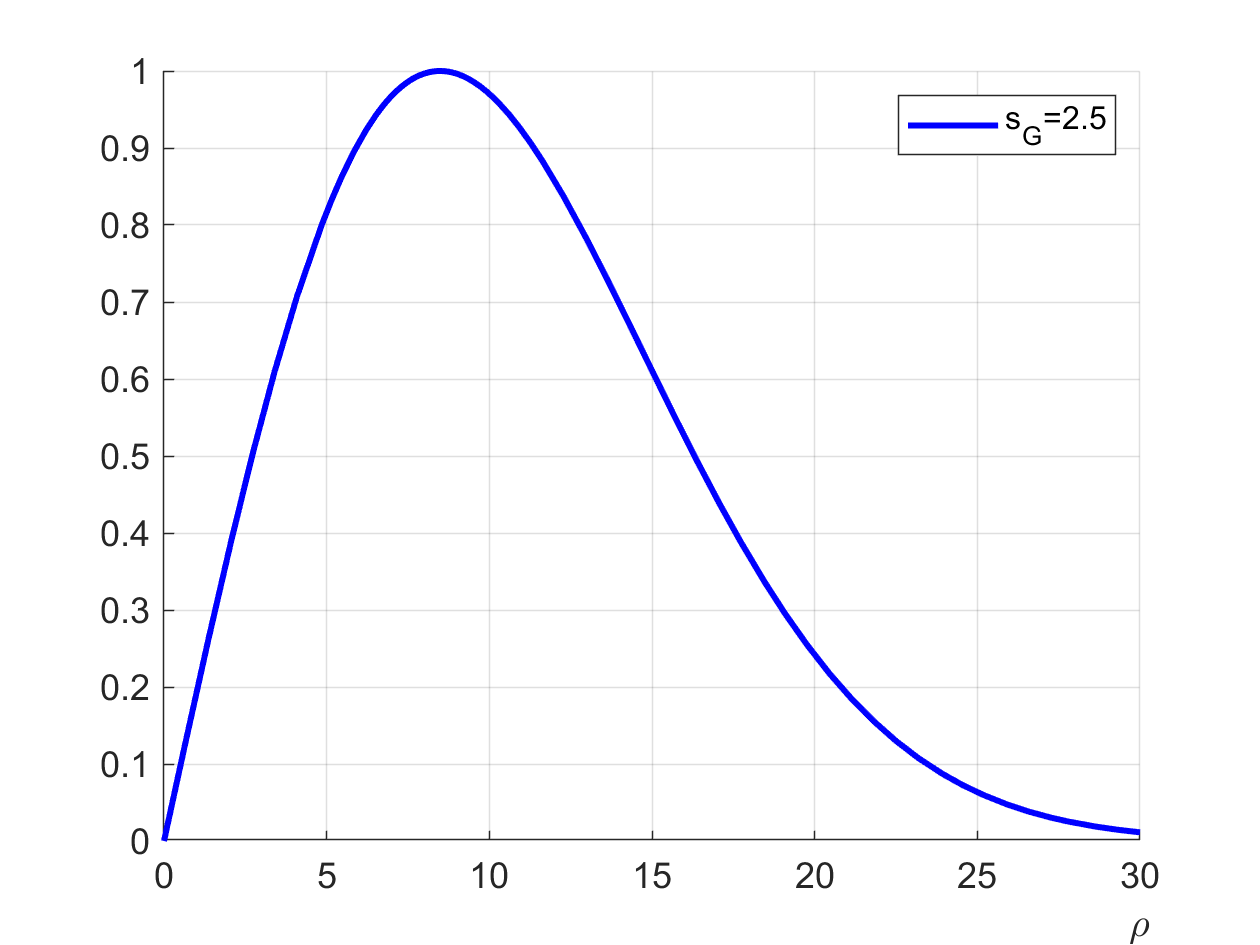}
\caption{The radial frequency response magnitude of the VTNF for $s_G=2.5$ \emph{arcmin} normalized with respect to its maximum value.}
\label{figVNTF}
\end{figure}

The magnitude of the VNTF frequency response versus the radial frequency is the same in any orientation. It is displayed in Fig.\ref{figVNTF}.

Here, and from now on, $s_G$ is assumed equal to $s_G=2.5$ \emph{arcmin}, unless otherwise noted. This choice sets the maximum of the radial frequency response magnitude at about 8.5 \emph{cycles/degree} in the radial frequency, according to the experimental data provided in \citep{CAMPBELL65} and \citep{WILLIAMS85}.

In the lowest spatial frequency range, the magnitude of the VNTF increases linearly. At higher spatial frequencies, the VNTF exhibits a soft-decaying low-pass behavior, reaching an attenuation of about 40 dB at 30 \emph{cycles/degree} at the Nyquist frequency.

This behavior can be interpreted by regarding the VTNF as the cascade of two basic operators:
\begin{itemize}
\item{
an orientation selective \emph{complex gradient} operator $\displaystyle\left(\frac{\partial}{\partial x_1}+j\frac{\partial}{\partial x_2}\right)$ \citep{REISERT08}, whose frequency response is obtained by the Fourier transform derivation rule:
\begin{equation}
j2\pi f_1+j(j2\pi f_2)=j(2\pi\rho cos{\vartheta}+j2\pi\rho sin{\vartheta})=j2\pi\rho e^{j\vartheta}\; ;
\label{eqn:derivtheorem}
\end{equation}
}
\item{
a radial frequency selective \emph{Gaussian smoothing} operator, represented by the frequency response:
\begin{equation}
G(\rho,\theta)=e^{-S_G^2\rho^2}
\label{eqn:gaussspectrum}
\end{equation}
which is responsible of a \emph{neural blur}.
}
\end{itemize}

Therefore, the visual map $y\left(\mathbf{p}\right)$ of the VRF is globally interpreted as a \emph{complex, Gaussian-smoothed gradient field} associated to the retinal image. The parameter $s_G$ will be referred to as the \emph{spread} of the VRF, or as the \emph{neural spread}.

Different from the \emph{complex Gabor} functions, whose paired real components are aligned each other \citep{DAUGMAN93,WATSON97}, the paired real components of the VRF are mutually orthogonal in the image plane (Fig.\ref{fig1}). As such, the VRF is \emph{steerable} \citep{SIMONCELLI92} and more specifically \emph{scalar steerable}, i.e., it rotates in azimuth by multiplication by a complex number.
\begin{equation}
\begin{array}{c}
h(r,\varphi-\alpha)=h(r,\varphi)e^{j\alpha}\; ; \\
H(\rho,\vartheta-\alpha)=H(\rho,\vartheta)e^{j\alpha}\; .
\end{array}
\label{eqn:VRFmodulatedfreq}
\end{equation}
This implies that rotated version of the VRFs by a generic azimuth $\alpha$ are obtained by  linear combinations of real components $Re{\left\{h(r,\varphi)\right\}}$ and $Im{\left\{h(r,\varphi)\right\}}$:
\begin{equation}
\begin{array}{c}
Re{\left\{h(r,\varphi-\alpha)\right\}}=Re{\left\{h(r,\varphi)\right\}}cos{\alpha}-Im{\left\{h(r,\varphi)\right\}}sin{\alpha}\; ; \\
Im{\left\{h(r,\varphi-\alpha)\right\}}=Re{\left\{h(r,\varphi)\right\}}sin{\alpha}+Im{\left\{h(r,\varphi)\right\}}cos{\alpha}\; .
\end{array}
\label{eqn:retinagridreim}
\end{equation}

\begin{figure}[!ht]
\centering
\includegraphics[width=3.1in]{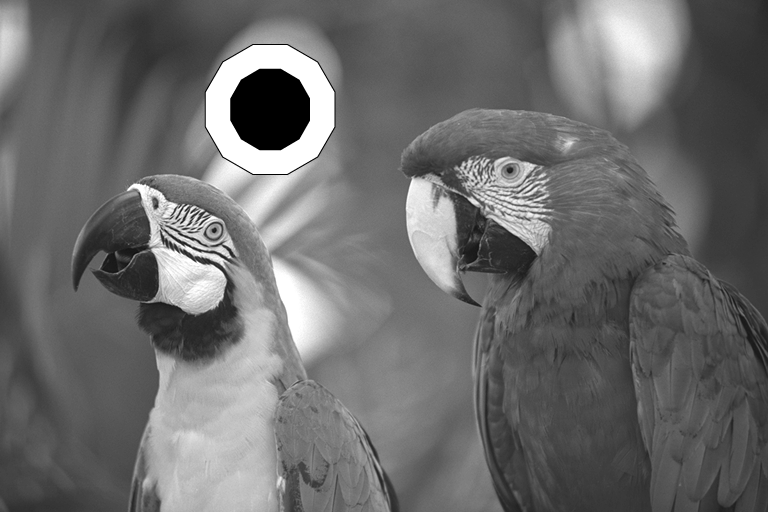}\vspace{3mm}
\includegraphics[width=3.1in]{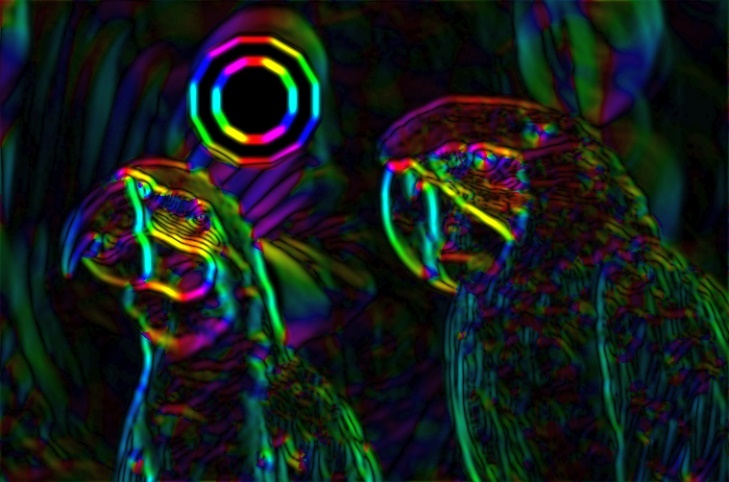}
\caption{The luminance of an image (left) of the database LIVE \citep{SHEIKH06B} compared to the corresponding visual map $y\left(\mathbf{p}\right)$ (right), where  the magnitude of the edges is coded into luminance, and their direction, in the interval $[0,\pi)$, into hue. The hue/direction code is read in the edges of the upper-left superimposed polygonal.}
\label{fig2}
\end{figure}

The VRF model materializes into the visual map $y\left(\mathbf{p}\right)$. In the correspondence of an \emph{edge} of $I\left(\mathbf{p}\right)$, the magnitude of $y\left(\mathbf{p}\right)$ measures the \emph{edge strength}, while the phase $tg^{-1}\left[\displaystyle\frac{Im{\left\{y\left(\mathbf{p}\right)\right\}}}{Re{\left\{y\left(\mathbf{p}\right)\right\}}}\right]$ indicates the orientation orthogonal to the edge \citep{CUSANI91,JACOVITTI95B,MCILHAGGA12}. In the example of Fig.\ref{fig2}, the luminance of a retinal image (left) and its visual map in false color (right) are displayed. In the visual map the gradient strength is indicated by the luminance component whereas, for visual immediateness, only the direction of the gradient in the interval $[0,\pi)$ is indicated with the hue color component.

The visual map $y\left(\mathbf{p}\right)$ is a \emph{near-complete}, \emph{sparse representation} of $I\left(\mathbf{p}\right)$. In fact, except for its mean value, $I\left(\mathbf{p}\right)$ can be fully recovered from $y\left(\mathbf{p}\right)$ by \emph{spectral inversion} (i.e., by division by $H(\rho,\vartheta)$ in the frequency domain).

Standard multichannel spatial vision models (see \citep{SCHUTT17} for a historical account) follow a tomographic-like approach. Around any point, they analyze the image from a limited number of azimuthal views. For each view, they apply co-oriented filters tuned to different bands. Their outputs are then combined in different ways.

In comparison, around any point, the VRF performs a full-band \emph{radial tomographic} analysis in every orientation, as described in \citep{CUSANI89,JACOVITTI90}. For the scope of this work, the outstanding advantage of the VRF is that its output (the visual map) allows straightforward computation of the PFI \citep{NERI04} as described in the next section.

The VRF is the simplest functional spatial vision model based on HAFs. It coincides with the first order component of the orthogonal family of the Laguerre Gauss (LG) functions \citep{VICTOR03,JACOVITTI00,MASSEY05} or, equivalently, of the 2D Hermite functions, which span the same signal space \citep{MARTENS90,DICLAUDIO11}. Higher order LG analysis provides \emph{functional} spatial vision models oriented to structures more complex that  simple edges \citep{DICLAUDIO10,NERI04}. HAF based \emph{wavelets} can be also used for multiresolution analysis \citep{JACOVITTI00}.

\section{The Positional Fisher Information}
\label{sec:The Positional Fisher Information}

A \emph{detail} $d_p(q)$ of a visual map $y\left(\mathbf{p}\right)$ is formally defined as:
\begin{equation}
d_\mathbf{p}(\mathbf{q})=w_\mathbf{p}(\mathbf{q}-\mathbf{p})\cdot y\left(\mathbf{q}-\mathbf{p}\right)
\label{eqn:detail}
\end{equation}
where $w_\mathbf{p}(\mathbf{q}-\mathbf{p})$ is a sampling window centered on $\mathbf{p}$.

A comprehensive calculus of the Fisher Information of a detail about its position, orientation, and scale, in the presence of a background Gaussian white noise, was provided in \citep{NERI04}. As specified in \citep{DICLAUDIO10}, the \emph{total} PFI of a detail is calculated as:
\begin{equation}
\psi(\mathbf{p})=\frac{\lambda(\mathbf{p})}{\sigma_V^2}
\label{eqn:PFI}
\end{equation}
where $\lambda(\mathbf{p})$ is the smoothed gradient energy of the detail, computed as
\begin{equation}
\lambda(\mathbf{p})=\sum_{\mathbf{q}}{w_\mathbf{p}(\mathbf{q})^2\left|y\left(\mathbf{q}-\mathbf{p}\right)\right|^2}
\label{eqn:lambda1}
\end{equation}
and $\sigma_V^2$ is the variance of the background noise\footnote{Strictly speaking, the PFI of a detail does not coincide with the PFI of the pattern “contained in” the detail. In fact, the window itself carries its own PFI. In the following, the latter contribution will be neglected, assuming that the window is so smooth that the information carried by its shape is small with respect to the information carried by the captured pattern.}.

The inverse square root of the PFI
\begin{equation}
e_{MIN}(\mathbf{p})=\sqrt{\frac{1}{\psi(\mathbf{p})}}=\sqrt{\frac{\sigma_V^2}{\lambda(\mathbf{p})}}
\label{eqn:emin}
\end{equation}
represents the minimum standard deviation of the detail position error $e_{MIN}(\mathbf{p})$, achievable with an unbiased estimator, irrespective of the employed estimation method \citep{TREES92}.
Thus, the quantity
\begin{equation}
\sqrt{\psi(\mathbf{p})}=\frac{1}{e_{MIN}(\mathbf{p})}
\label{eqn:sqrtpsi}
\end{equation}
measures the \emph{certainty} of the detail position in the visual plane. For a given amount of background noise, the higher the smoothed gradient energy, the higher the PFI, the greater the certainty about the detail position.

The smoothed gradient energy $\lambda(\mathbf{p})$ is significantly expressed by the (two-dimensional) Fourier transform of the detail. Applying the Parseval theorem (which equates the energy calculated in the space and in the spatial frequency domains) the PFI of the detail represented by its Fourier transform $D_\mathbf{p}(\rho,\vartheta)$ is
\begin{equation}
\psi(\mathbf{p})=\frac{1}{\sigma_V^2}\int_{0}^{2\pi}\int_{0}^{+\infty}{4\pi^2\rho^2\left|G(\rho,\vartheta)\right|^2\left|D_\mathbf{p}(\rho,\vartheta)\right|^2\left|B(\rho,\vartheta)\right|^2\rho d\rho d\vartheta}
\label{eqn:psi}
\end{equation}
where $B(\rho,\vartheta)$ is the Optical Transfer Function (OTF), i.e., the spatial frequency response of the optical system, from the observed object to the retina \citep{WATSON13}. Its inverse 2D transform is the Optical PSF $b(r,\varphi)$.

The overall OTF of a vision system, including the human eye, is a combination of the OTFs of cascaded subsystems including the OTF of correcting lenses, the OTF of an imaging system, the OTF of a display system, etc..

Under the hypothesis of linearity, the overall OTF is the product of the single OTFs. In other terms, the overall Optical PSF is the cascaded 2D convolution of the single PSFs.

\section{The natural scene spectrum and the PFI equivalence}
\label{sec:The natural scene spectrum and the PFI equivalence}

\subsection{The PFI of natural scenes}
\label{sec:The PFI of natural scenes}

For a generic natural image, the average PFI calculated on a group of N details visited during the visual exploration is:
\begin{equation}
\begin{array}{c}
\displaystyle\frac{\sum_{\mathbf{p}}{\lambda(\mathbf{p})}}{N\sigma_V^2}= \\
\displaystyle\frac{1}{\sigma_V^2}\int_{0}^{2\pi}\int_{0}^{+\infty}{4\pi^2\rho^2\left|G(\rho,\vartheta)\right|^2\left|D_N(\rho,\vartheta)\right|^2\left|B(\rho,\vartheta)\right|^2\rho d\rho d\vartheta}
\end{array}
\label{eqn:PFINdetails}
\end{equation}
where
\begin{equation}
\left|D_N(\rho,\vartheta)\right|^2=\frac{1}{N}{\sum_{\mathbf{p}}\left|D_\mathbf{p}(\rho,\vartheta)\right|}^2
\label{eqn:AED}
\end{equation}
is the average energy spectrum of the $N$ visited details.
The expected value $\Psi$ of the PFI over the random set of natural images is defined as:
\begin{equation}
\begin{array}{c}
\Psi\doteq E\left\{\displaystyle\frac{\sum_{\mathbf{p}}{\lambda(\mathbf{p})}}{N\sigma_V^2}\right\}= \\
\displaystyle\frac{1}{\sigma_V^2}\int_{0}^{2\pi}\int_{0}^{+\infty}{4\pi^2\rho^2\left|G(\rho,\vartheta)\right|^2E\left\{\left|D_N(\rho,\vartheta)\right|^2\right\}\left|B(\rho,\vartheta)\right|^2\rho d\rho d\vartheta}
\end{array}
\label{eqn:Psi}
\end{equation}
where $E\left\{\left|D_N(\rho,\vartheta)\right|^2\right\}$ denotes the expected value of the energy spectrum of the visited details over the random set of natural images.

It is well known that the expected value of the energy spectrum of natural images is proportional to the inverse of the square of the radial frequency $\displaystyle\frac{1}{\rho^2}$. The generality of this spectral distribution is supported by theoretical arguments \citep{SIMONCELLI01,FIELD97,GRAHAM06,KUANG12,SHAAF96,BELL97}.

Here, this property is attributed to the \emph{average spectrum} of the \emph{visited details}. Posing, for the sake of generality \citep{TORRALBA03}
\begin{equation}
E\left\{\left|D_N(\rho,\vartheta)\right|^2\right\}=f(\vartheta)\frac{1}{\rho^2}\; ;
\label{eqn:EAED}
\end{equation}
and absorbing $4\pi^2$ in $f(\vartheta)$ it follows that
\begin{equation}
\Psi=\frac{1}{\sigma_V^2}\int_{0}^{2\pi}{f\left(\vartheta\right)\int_{0}^{+\infty}{\left|G(\rho,\vartheta)\right|^2\left|B(\rho,\vartheta)\right|^2\rho d\rho d\vartheta}}
\label{eqn:PSINdetails}
\end{equation}
and, in the absence of blur:
\begin{equation}
\Psi_0=\frac{1}{\sigma_V^2}\int_{0}^{2\pi}{f\left(\vartheta\right)\int_{0}^{+\infty}{\left|G(\rho,\vartheta)\right|^2\rho d\rho d\vartheta}}\; .
\label{eqn:PSI0}
\end{equation}
In the case of isotropic Gaussian blur, the OTF is
\begin{equation}
B(\rho,\vartheta)=e^{-s_B^2\rho^2}
\label{eqn:OTF}
\end{equation}
where $s_B$ is referred to as the \emph{optical spread}. Therefore,
\begin{equation}
\Psi=\frac{F}{\sigma_V^2}\int_{0}^{+\infty}{e^{-2(s_G^2+s_B^2)\rho^2}\rho d\rho}
\label{eqn:PSINdetails1}
\end{equation}
\begin{equation}
\Psi_0=\frac{F}{\sigma_V^2}\int_{0}^{+\infty}{e^{-2s_G^2\rho^2}\rho d\rho}
\label{eqn:PSI01}
\end{equation}
where the coefficient $F$ is
\begin{equation}
F=\int_{0}^{2\pi}f\left(\vartheta\right)d\vartheta\; .
\label{eqn:Fcoeff}
\end{equation}
Finally, from the equality
\begin{equation}
\int_{0}^{+\infty}{e^{-2(s_G^2+s_B^2)\rho^2}\rho d\rho=\frac{1}{4(s_G^2+s_B^2)}}
\label{eqn:intsolution}
\end{equation}
it follows that
\begin{equation}
\frac{\Psi}{\Psi_0}=\frac{s_G^2}{s_G^2+s_B^2}\; .
\label{eqn:PSIratio}
\end{equation}

\subsection{The PFI equivalence}
\label{sec:The PFI equivalence}

\begin{figure}[!htb]
\centering
\includegraphics[width=3.1in]{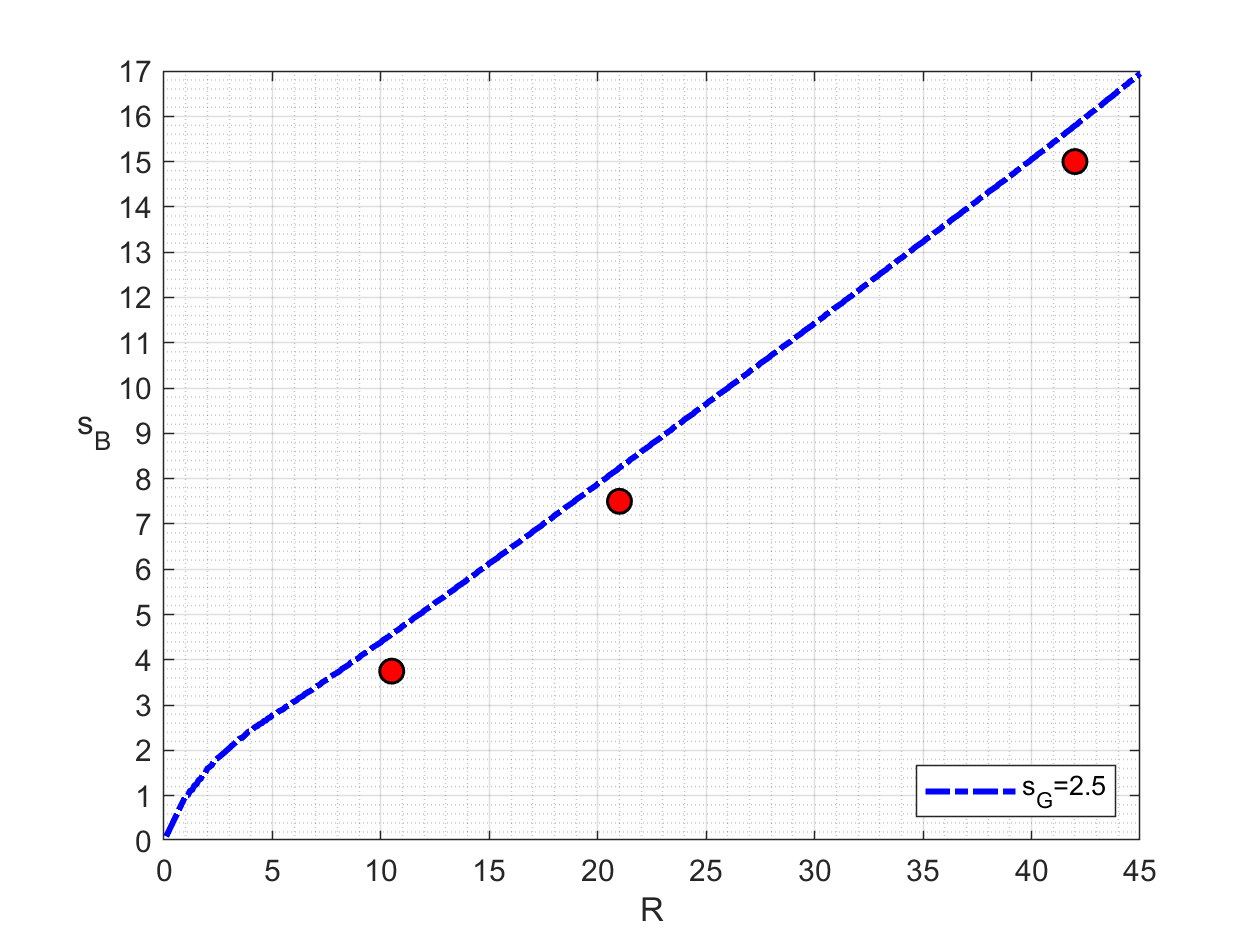}
\caption{The radius of the sinc blur versus the spread of its PFI equivalent Gaussian blur. The red points indicate some average results of experiments conducted with synthetic images emulating natural images \citep{MURRAY10}.}
\label{figsBR}
\end{figure}

\begin{figure*}[!ht]
\centering
\includegraphics[width=3.1in]{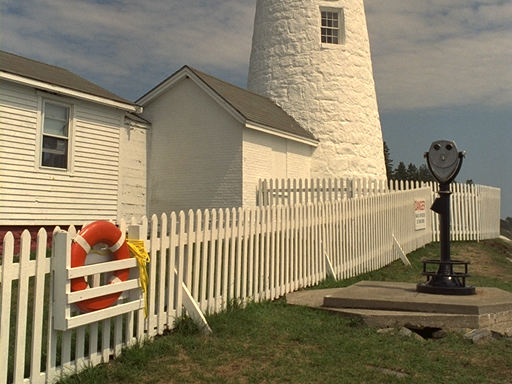}\hspace{3mm}
\includegraphics[width=3.1in]{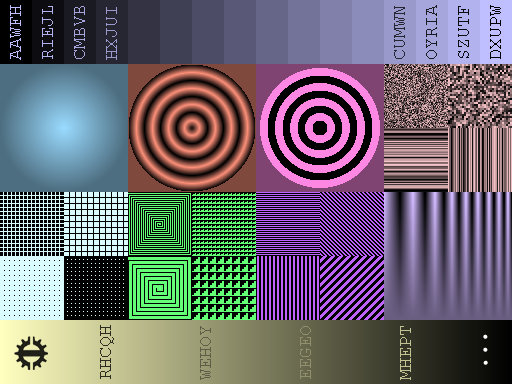}\vspace{3mm}
\includegraphics[width=3.1in]{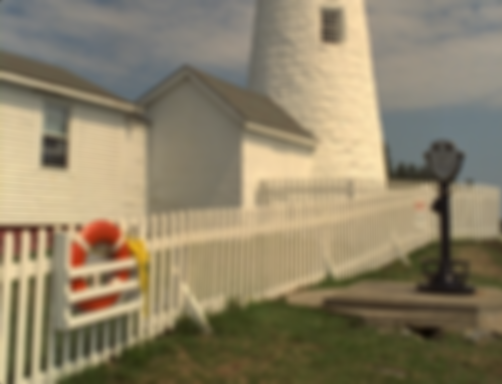}\hspace{3mm}
\includegraphics[width=3.1in]{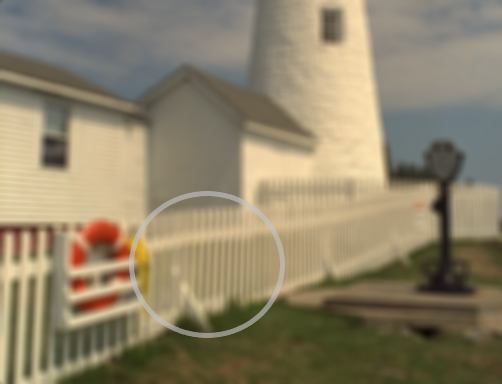}\vspace{3mm}
\includegraphics[width=3.1in]{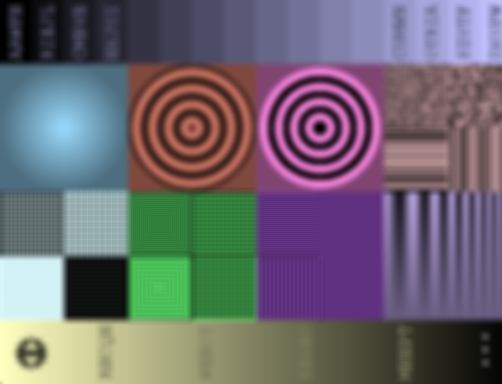}\hspace{3mm}
\includegraphics[width=3.1in]{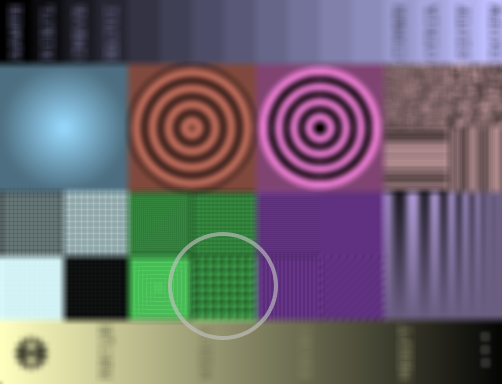}
\caption{Upper row: two original images, Second and third rows: Gaussian blurred image (left) and Fisher equivalent sinc blurred images (right). The values of $R$ were chosen to put into evidence the effects of the spectral sidelobes of the sinc blur (see the encircled patterns). Natural image (i19 of the database): $R=6$ \emph{arcmin}, corresponding to $s_B=3.1$ \emph{arcmin}. Synthetic image (i25 of the database): $R=7$ \emph{arcmin}, corresponding to $s_B=3.4$ \emph{arcmin}.}
\label{fig3}
\end{figure*}

\begin{figure*}[!ht]
\centering
\includegraphics[width=3.1in]{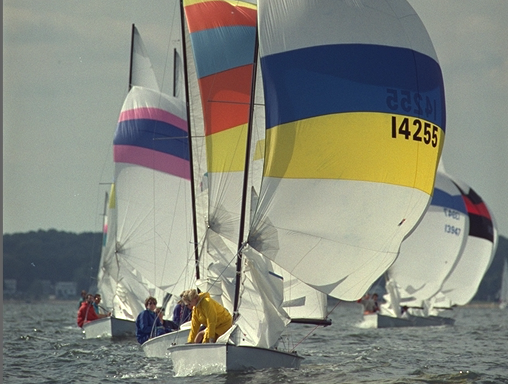}\hspace{3mm}
\includegraphics[width=3.1in]{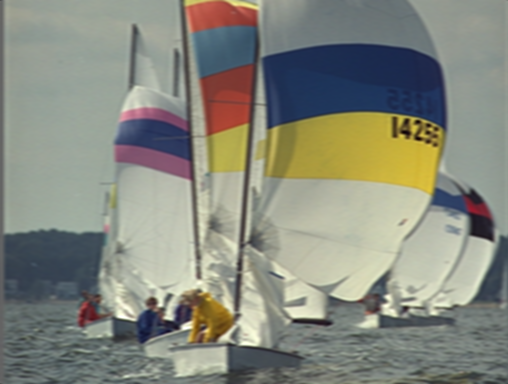}\vspace{3mm}
\includegraphics[width=3.1in]{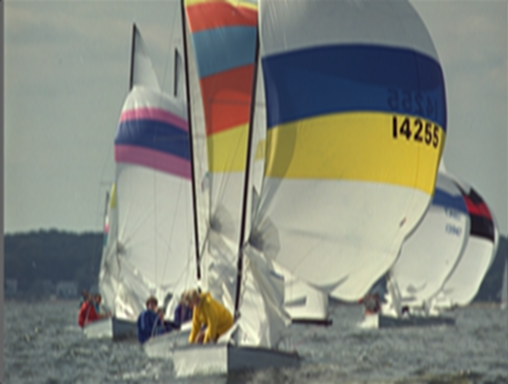}\hspace{3mm}
\includegraphics[width=3.1in]{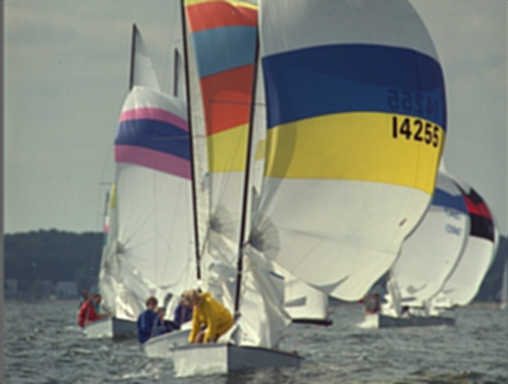}
\caption{A unblurred image (upper left) and its astigmatic blurred versions. For $s_H=4$, $s_V=1$ (horizontal blur, upper right), and for $s_H=1$, $s_V=4$ (vertical blur, lower left). They are PFI equivalent to the isotropic Gassian blurred version shown in the lower right image with $s_B=2.54$.}
\label{fig4}
\end{figure*}

The above result is so simple owing to the Gaussian shape of the blur. However, having assumed that the blur discomfort depends only on the PFI loss, the discomfort due to different types of blur could be predicted by the same formula applying the concept of \emph{PFI equivalence}.

\theoremstyle{definition}
\newtheorem{definition}{Definition}[section]
\begin{definition}[PFI Equivalence]
A blur characterized by the generic OTF $B(\rho,\vartheta)$ is said to be PFI equivalent to an isotropic Gaussian blur with standard deviation (spread) $s_B$ if it yields the same expected PFI for natural images, i.e.:
\begin{equation}
\begin{array}{c}
\displaystyle\int_{0}^{2\pi}\int_{0}^{+\infty}{e^{-2s_G^2\rho^2}\left|B(\rho,\vartheta))\right|^2\rho d\rho d\vartheta}= \\
\displaystyle\int_{0}^{2\pi}\int_{0}^{+\infty}{e^{-2(s_G^2+{s^\prime}_B^2)\rho^2}\rho d\rho d\vartheta}\; .
\end{array}
\label{eqn:intequiv}
\end{equation}
\end{definition}

This equivalence criterion is intuitive. It equals the energies of the actual OTF and of an isotropic Gaussian OTF, both weighted by the squared magnitude of the VNTF. In particular, the PFI equivalence does not depend on the phase of the OTF.

An important  example of PFI equivalence is the one of the out-of-focus blur, whose PSF is modeled as a cylinder of unitary volume and radius $R$. It is referred also to as the \emph{disc blur}, or the \emph{sinc blur}, and is characterized by the following OTF:
\begin{equation}
B(\rho,\vartheta)=2\frac{J_1(2\pi\rho R)}{2\pi\rho R}\; .
\label{eqn:Besselsinc}
\end{equation}
Equating the PFI of the sinc blur and of the Gaussian isotropic blur yields
\begin{equation}
\begin{array}{c}
\displaystyle4\int_{0}^{+\infty}{\left(\frac{J_1(2\pi\rho R)}{2\pi\rho R}\right)^2e^{-2s_G^2\rho^2}\rho d\rho}= \\
\displaystyle\int_{0}^{+\infty}{e^{-2(s_G^2+s_B^2)\rho^2}\rho d\rho}=\frac{1}{4(s_G^2+s_B^2)}\; .
\end{array}
\label{eqn:intequiv1}
\end{equation}

The left side integral is not available in closed form. A careful numerical integration provides a value of the optical spread $s_B$ of the isotropic Gaussian blur as a function of its PFI equivalent sinc blur of radius $R$, as plotted in Fig.\ref{figsBR} and is roughly expressed by the rule $\displaystyle\frac{s_B}{R}\approx\frac{3}{8}$. In the same figure some subjective equivalence judgments averaged over a pool of six observers are also reported. These empirical data refer to synthetic images whose contours emulate the ones of natural images \citep{MURRAY10}.

%\begin{figure}[!htb]
%\centering
%\includegraphics[width=3.1in]{R-sB.png}
%\caption{The radius of the sinc blur versus the spread of its PFI equivalent Gaussian blur. The red points indicate some average results of experiments conducted with synthetic images emulating natural images \citep{MURRAY10}.}
%\label{figsBR}
%\end{figure}

From a perceptual viewpoint, it appears that this equivalence works generally well. Here, to provide the reader with visual examples, the most critical cases in the Tampere image database (TID2013) \citep{PONOMARENKO13}, are reported in Fig.\ref{fig3}, including the non-natural image i25 as a benchmark. Notice that some grating patterns are cancelled out or amplified by the sidelobes of the OTF of the sinc blur in comparison to the Gaussian blur. Notice also that these effects are not present in the natural out-of-focus blur, owing to the apodization of the pupil, which attenuates the sidelobes \citep{ZHANG99}.

%\begin{figure*}[!ht]
%\centering
%\includegraphics[width=3.1in]{Original_I19.png}\hspace{3mm}
%\includegraphics[width=3.1in]{Original_i25_mod.png}\vspace{3mm}
%\includegraphics[width=3.1in]{Gaussian_Blur_ifft_I19_sB3p1.png}\hspace{3mm}
%\includegraphics[width=3.1in]{Bessel1_Blur_ifft_I19_R6.png}\vspace{3mm}
%\includegraphics[width=3.1in]{Gaussian_Blur_ifft_i25_mod_sB3p4.png}\hspace{3mm}
%\includegraphics[width=3.1in]{Bessel1_Blur_ifft_i25_mod_R7.png}
%\caption{Upper row: two original images, Second and third rows: Gaussian blurred image (left) and Fisher equivalent sinc blurred images (right). The values of $R$ were chosen to put into evidence the effects of the spectral sidelobes of the sinc blur (see the encircled patterns). Natural image (i19 of the database): $R=6$ \emph{arcmin}, corresponding to $s_B=3.1$ \emph{arcmin}. Synthetic image (i25 of the database): $R=7$ \emph{arcmin}, corresponding to $s_B=3.4$ \emph{arcmin}.}
%\label{fig3}
%\end{figure*}

A second example of PFI equivalence regards the non-isotropic Gaussian blur, referred to as \emph{astigmatic} Gaussian blur. For the sake of simplicity, this equivalence is calculated here only for the case of isotropic image spectral energy distribution. Using for convenience the Cartesian coordinates, the OTF of this blur is:
\begin{equation}
B(f_1,f_2)=e^{-2(s_V^2f_1^2+s_H^2f_2^2)}
\label{eqn:OTFfreq}
\end{equation}
where $s_H$ and $s_V$ are the horizontal and vertical optical spreads. A straightforward algebraic analysis shows that the astigmatic Gaussian blur is PFI equivalent to the isotropic Gaussian blur with spread
\begin{equation}
s_B=\sqrt{\sqrt{s_G^4+s_G^2(s_H^2+s_V^2)+s_V^2s_H^2}-s_G^2}\; .
\label{eqn:sBisotropic}
\end{equation}

An example is provided in Fig.\ref{fig4}. Two versions of an original image, respectively affected by an astigmatic Gaussian blur with $s_H=4$, $s_V=1$ and with $s_H=1$, $s_V=4$ are compared to the same image affected by their PFI equivalent isotropic Gaussian blur. Notice that the sea waves are better localized in presence of horizontal blur, while masts are better localized in presence of vertical blur. The average localizability loss is visually balanced by the isotropic Gaussian blur.

%\begin{figure*}[!ht]
%\centering
%\includegraphics[width=3.1in]{Original_I09.png}\hspace{3mm}
%\includegraphics[width=3.1in]{GaussianAstigmatic_Blur_ifft_I09_sH4_sV1.png}\vspace{3mm}
%\includegraphics[width=3.1in]{GaussianAstigmatic_Blur_ifft_I09_sH1_sV4.png}\hspace{3mm}
%\includegraphics[width=3.1in]{Gaussian_Blur_ifft_I09_sB2p54.png}
%\caption{A unblurred image (upper left) and its astigmatic blurred versions. For $s_H=4$, $s_V=1$ (horizontal blur, upper right), and for $s_H=1$, $s_V=4$ (vertical blur, lower left). They are PFI equivalent to the isotropic Gassian blurred version shown in the lower right image with $s_B=2.54$.}
%\label{fig4}
%\end{figure*}

\section{Measuring the blur discomfort}
\label{sec:Measuring the blur discomfort}

Provisionally, the amount of blur discomfort is assumed \emph{proportional} to the \emph{relative certainty loss} about the details position\footnote{An exponentiated version of such a measure was employed in \citep{DICLAUDIO18}.} defined as:
\begin{equation}
\varepsilon=\frac{\sqrt{\Psi_0}-\sqrt\Psi}{\sqrt{\Psi_0}}=1-\sqrt{\frac{\Psi}{\Psi_0}}\; .
\label{eqn:epsilon}
\end{equation}

This assumption is suggested by the belief that the perceived cost of wrong localization is proportional to the uncertainty of the Euclidean distances, at least for small errors.

Then, for the random set of natural images and for isotropic Gaussian blur, using \eqref{eqn:PSIratio} the relative certainty loss takes the form of the following a-dimensional \emph{discomfort index}:
\begin{equation}
\varepsilon=1-\sqrt{\frac{\Psi}{\Psi_0}}=1-\sqrt{\frac{1}{1+\left(\displaystyle\frac{s_B}{s_G}\right)^2}}\; .
\label{eqn:epsilon1}
\end{equation}

This index ranges between 0 (in the absence of blur) to 1 (for diverging blur). It depends only on the \emph{normalized blur}, i.e., the ratio between the optical spread on the retinal image and the neural spread of the VRF.

The optical blur spread $s_B$ is subject to change by the action of the natural accommodation system and, in the context of a composite optical system, by the action of technical devices. The neural spread $s_G$ plays instead the role of an inner reference. At glance, it could be argued that it is a stable parameter. However, some experiments indicate that the $s_G$ value is adaptive \citep{WEBSTER02}. It appears that the  visual adaption to a blurred image causes a dilation of the spread $s_G$, leading to a reduction of the normalized blur, so that the spectrum of the image looks wider and the image sharper.

The discomfort formula is now applied to the blur caused by out-of-focus condition of the eye optics, i.e., in \emph{natural vision}. Using the geometrical arguments of \citep{STRASBURGER18}, it is deduced that the radius $R$ of the out-of-focus optical PSF in arcmin is
\begin{equation}
R=1.71p\left|D\right|
\label{eqn:R}
\end{equation}
where $p$ is the pupil diameter (in mm) and the $D$ is the out-of-focus measure in diopter units ($m^{-1}$).

Applying the $\frac{s_B}{R}\approx\frac{3}{8}$ rule it follows that the spread of the PFI equivalent blur of the out-of-focus blur measured in diopters is approximated as
\begin{equation}
s_B\approx0.64p\left|D\right|
\label{eqn:sB1}
\end{equation}
so that a coarse estimate of the blur discomfort in natural vision is:
\begin{equation}
\varepsilon=1-\sqrt{\frac{1}{1+\left(0.64\displaystyle\frac{p}{s_G}D\right)^2}}\; .
\label{eqn:epsilon2}
\end{equation}

\begin{figure}[!ht]
\centering
\includegraphics[width=3.1in]{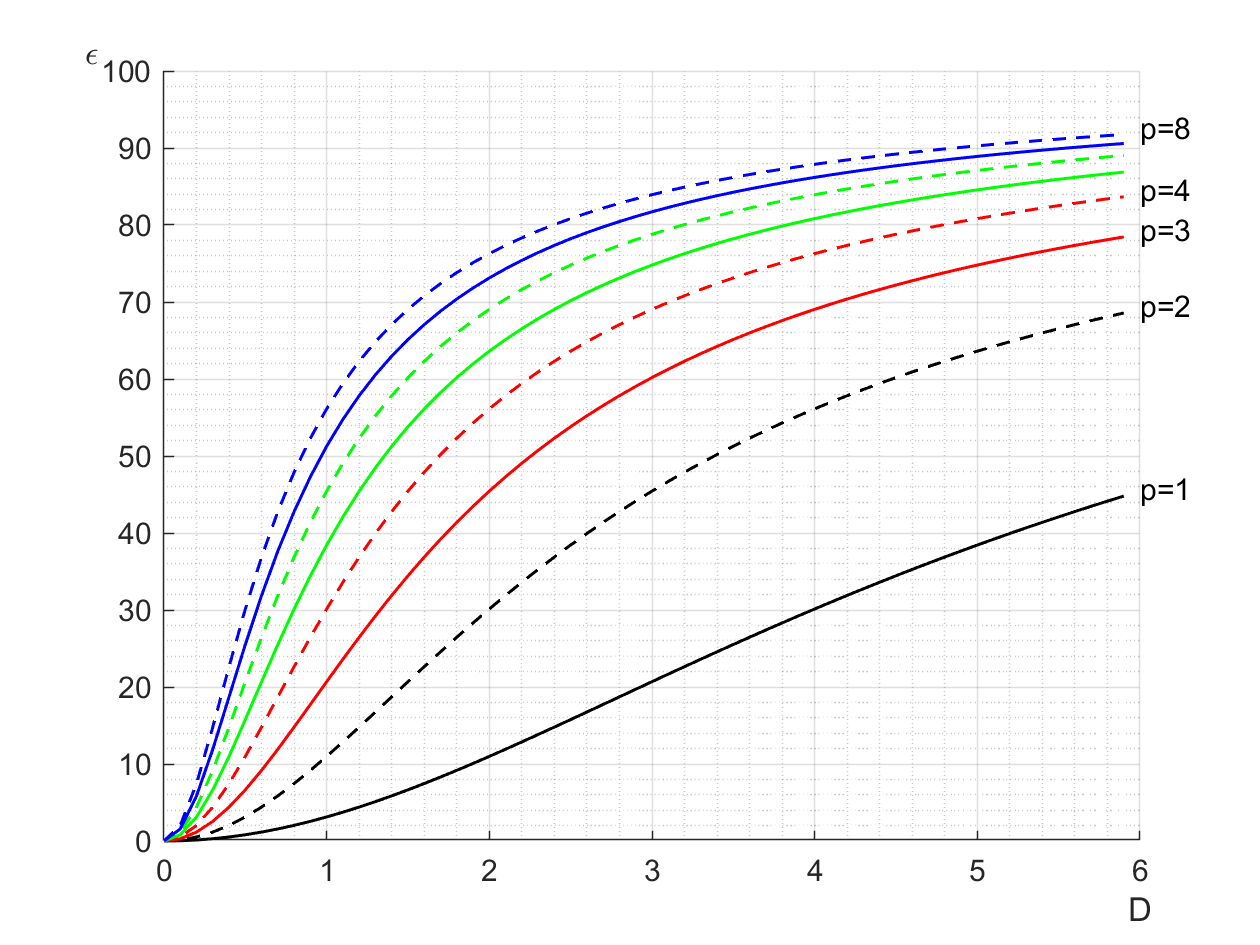}
\caption{The value of the theoretical discomfort index in centesimal units versus the diopters (measured in $m^{-1}$) for different values of the pupil diameter (measured in $mm$). (A typical pupil diameter when reading at normal illumination is 3 mm).}
\label{fig5}
\end{figure}

In the chart of Fig.\ref{fig5} this theoretical discomfort index $\varepsilon$ is plotted versus $D$ for different pupil diameters. The scale of $\varepsilon$ is expressed in centesimal units.

This chart essentially says that, if the out-of-focus discomfort is proportional to the relative certainty loss, as assumed, it is not linear with the blur spread. Moreover, since the sensitivity of the discomfort index with respect to the normalized spread $\xi\doteq\frac{s_B}{s_G}$ is calculated as
\begin{equation}
\frac{d\varepsilon}{d\xi}=\xi\left[\frac{1}{1+\xi^2}\right]^\frac{3}{2}\; ;
\label{eqn:deps_sBsG}
\end{equation}
the increment $\Delta\xi$ necessary to produce a given increment $\Delta\varepsilon$ is approximated as:
\begin{equation}
\Delta\xi=\frac{1}{\xi}\left[1+\xi^2\right]^\frac{3}{2}\Delta\varepsilon\; .
\label{eqn:deltasBsG}
\end{equation}

\begin{figure}[!ht]
\centering
\includegraphics[width=3.1in]{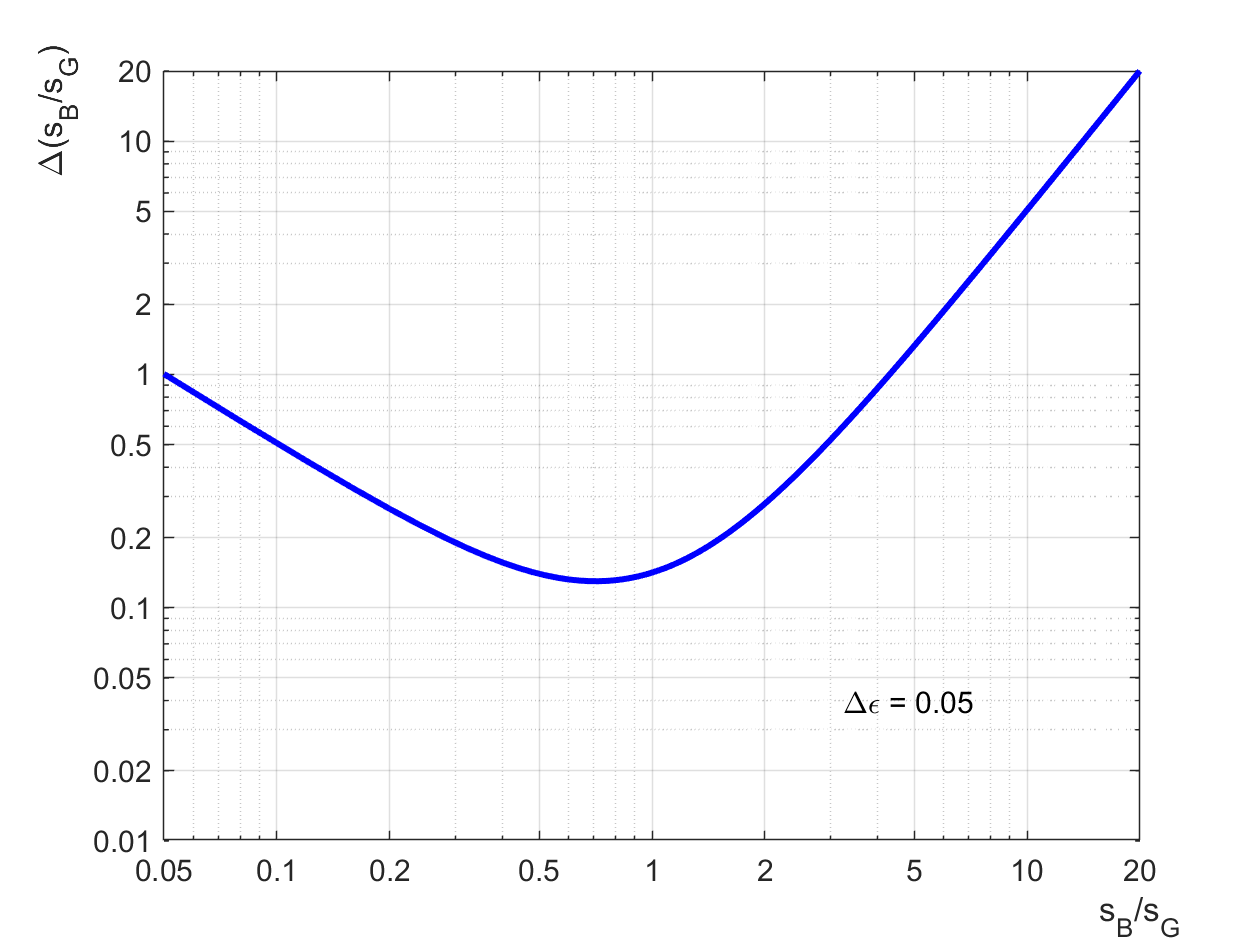}
\caption{The theoretical increment $\Delta\left(\frac{s_B}{s_G}\right)$ versus $\frac{s_B}{s_G}$ for $\Delta\varepsilon=0.05$ in a log/log scale. Curves for different values of $\Delta\varepsilon$ are obtained by vertical translation.}
\label{fig6}
\end{figure}

This increment exhibits a typical “dipper shape” \citep{OHARE11,SOLOMON09} as shown in Fig.\ref{fig6}. The theoretical minimum occurs in the correspondence of the normalized spread value $\xi=\displaystyle\frac{s_B}{s_G}=\frac{1}{\sqrt2}$.

In the above formulas, diffraction and aberration contributions to blur discomfort are considered negligible. Non-negligible aberrations could be accounted for by adopting parametric models for characterizing their PSFs \citep{GOODMAN05,WATSON15}, and calculating the PFI equivalent blur. A simple example is the aberration due to astigmatic Gaussian blur provided in Sect.4, where the blur is characterized by the two parameters $s_H$ and $s_V$. However, broad generalizations would be quite an undertaking beyond the scope of this paper.

\section{Experimental verification}
\label{sec:Experimental verification}

\begin{figure*}[!ht]
\centering
\includegraphics[width=3.1in]{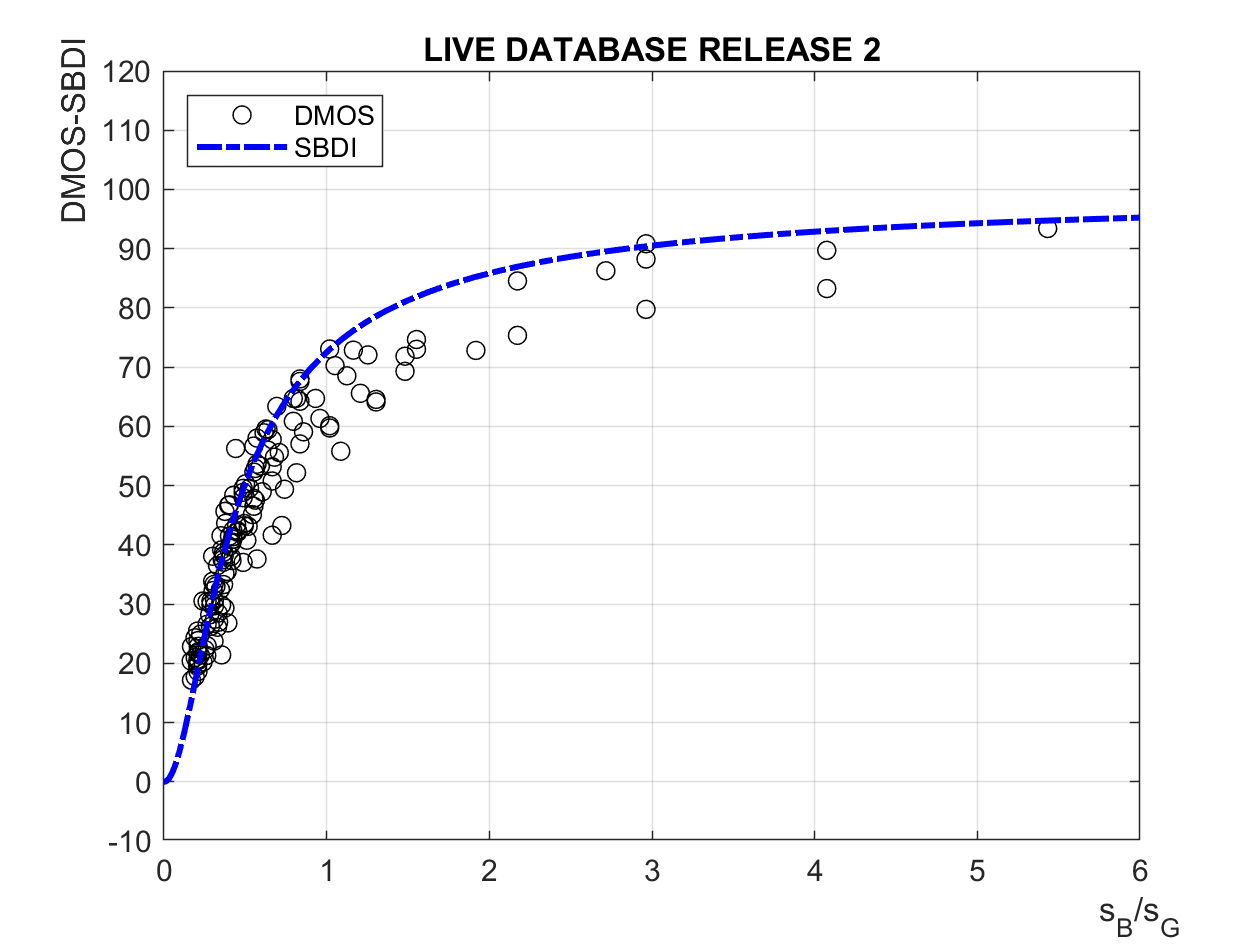}
\includegraphics[width=3.1in]{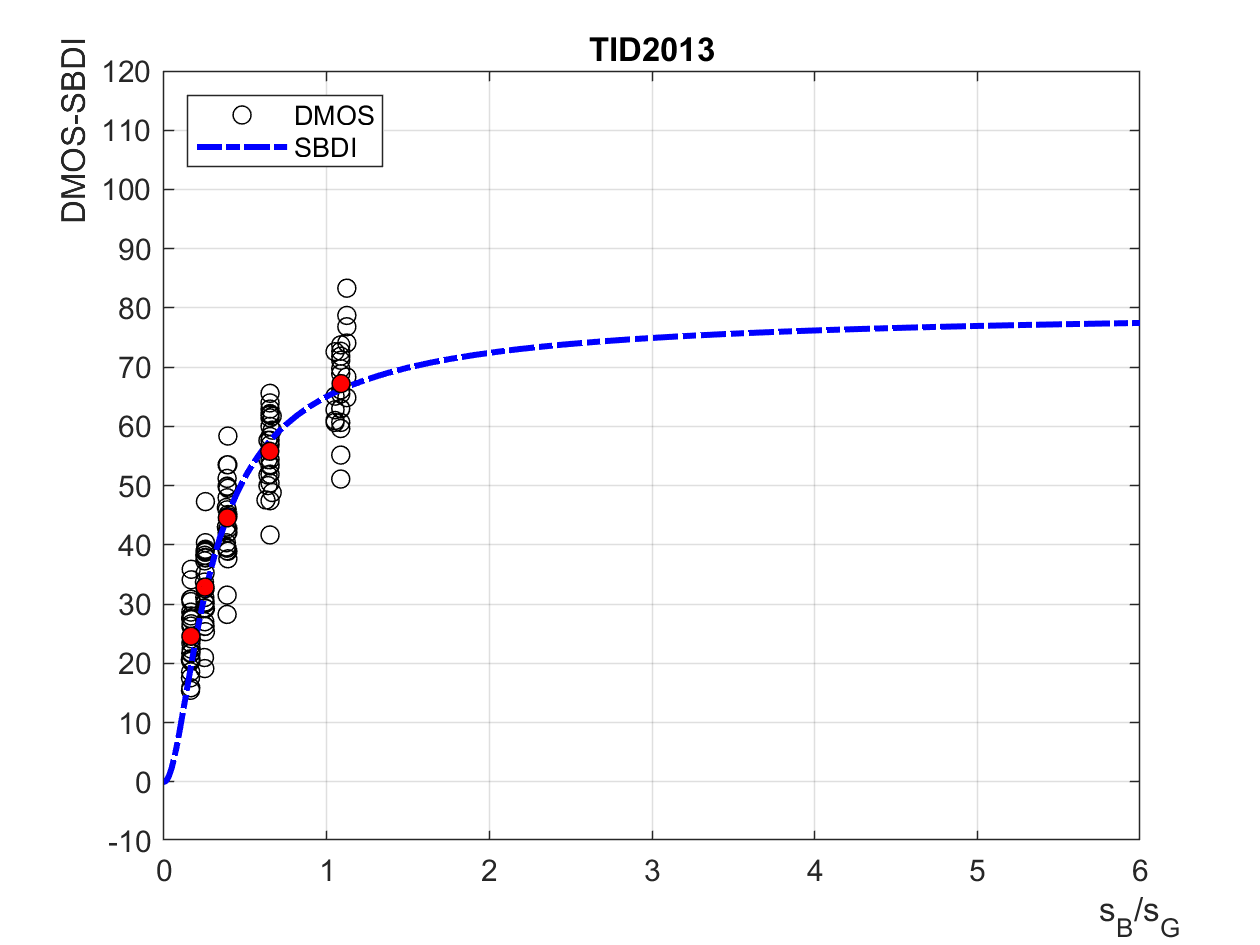}
\includegraphics[width=3.1in]{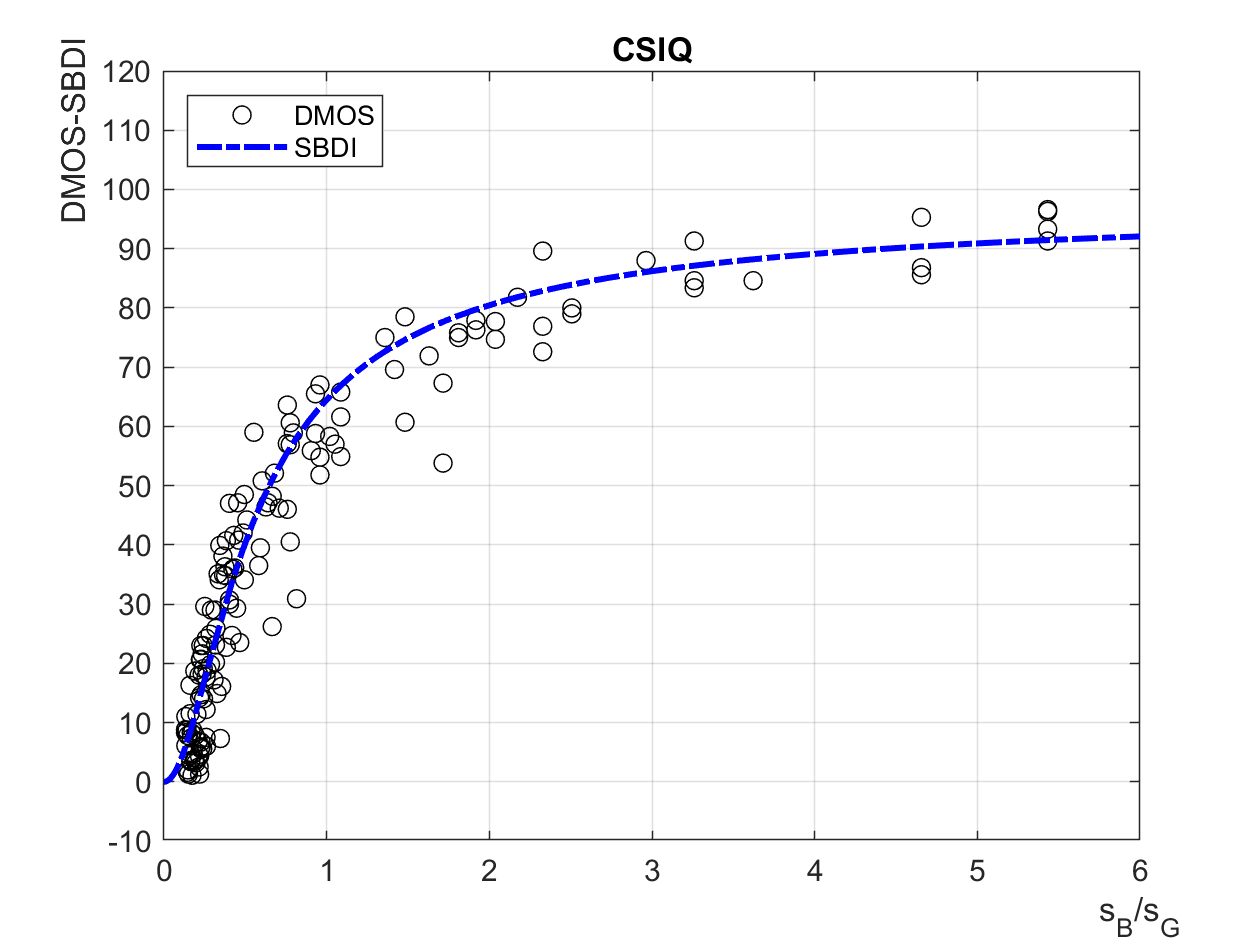}
\includegraphics[width=3.1in]{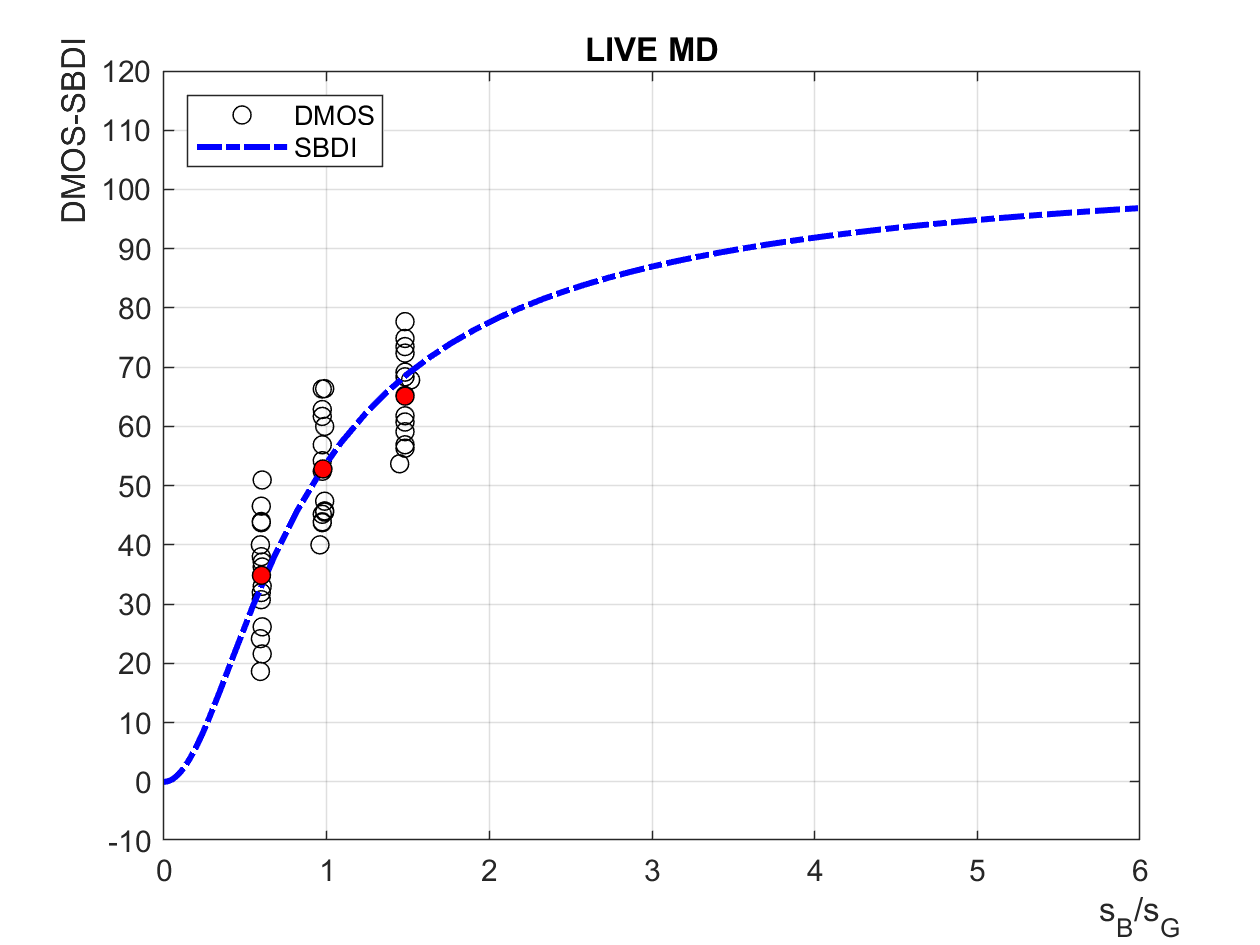}
\caption{The empirical DMOS values of the subjective quality loss for the blurred images of the databases versus the normalized blur spread compared to the values predicted by the theoretical model (dashed curves). Averages with respect to images are indicated by the filled circles in the TID2013 and LIVE MD scatterplots.}
\label{fig7}
\end{figure*}

The above results about the phenomenon of blur discomfort for natural images are derived by principles and assumptions. To assess their effectiveness in the reality, the theoretical model was first verified against empirical ratings of the \emph{subjective quality loss} of images caused by blur, which are argued to be strictly related to the blur discomfort. These data were released in response to the growing demand by the media industry for reliable automatic image quality assessment (IQA) through objective metrics. Four independent IQA databases were employed, based on different methodologies and protocols, and following different strategies to prevent biases and side effects.

IQA databases include images affected by Gaussian blur, which is considered sufficiently representative of the perceptual effect of the blur in many technical applications.

Subsequently, the predictions of the model were compared to experimental data where subjects were literally asked to rate “the \emph{visual discomfort}” due to blur \citep{OHARE13}.

All these experiments do not account for blur discomfort secondary effects. Furthermore, in these experiments the eye optical blur of the observers is corrected if present. So, the blur applied to the observed images emulates an undesired natural optical blur on the retina plane. In fact, under the hypothesis of linearity, the natural optical blur and the artificial blur applied to the observed images are interchangeable, because of the well-known property of cascaded convolutions.

Before presenting the experiments and discussing the results, it is essential to illustrate how the discomfort index must be scaled to fit the experimental settings of a database.

\subsection{The Scaled Blur Discomfort Index}
\label{sec:The Scaled Blur Discomfort Index}

\begin{figure*}[!ht]
\centering
\includegraphics[width=3.1in]{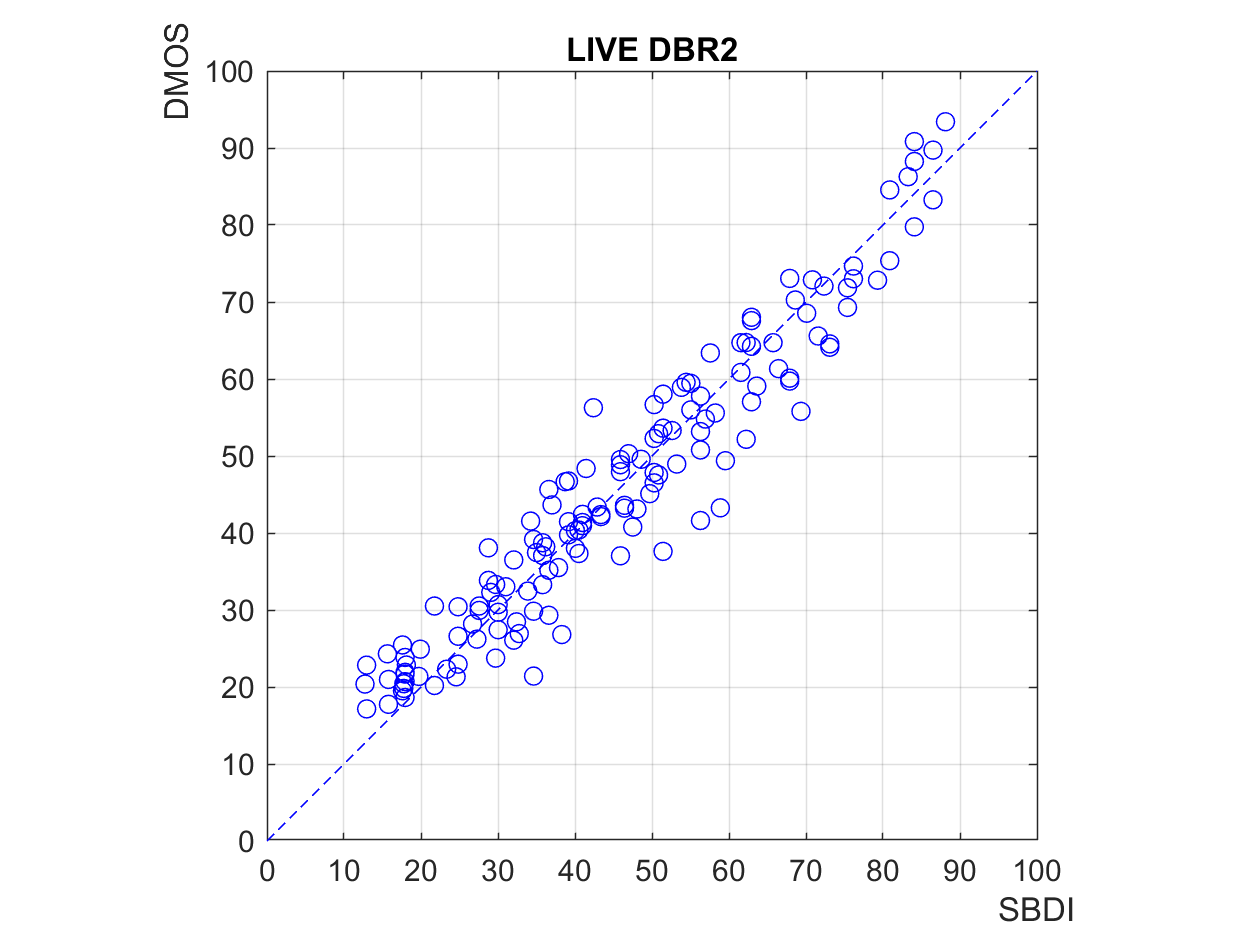}
\includegraphics[width=3.1in]{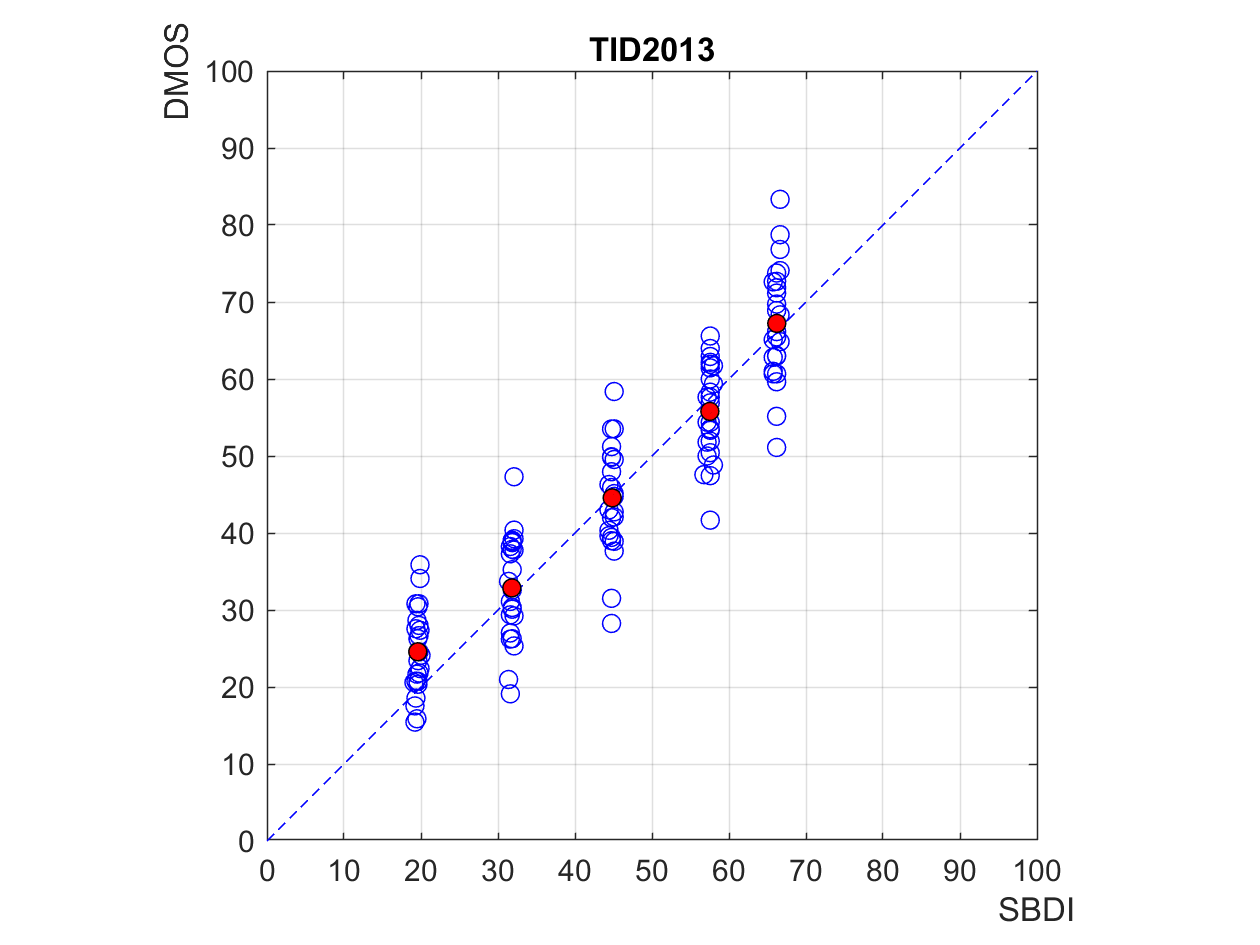}
\includegraphics[width=3.1in]{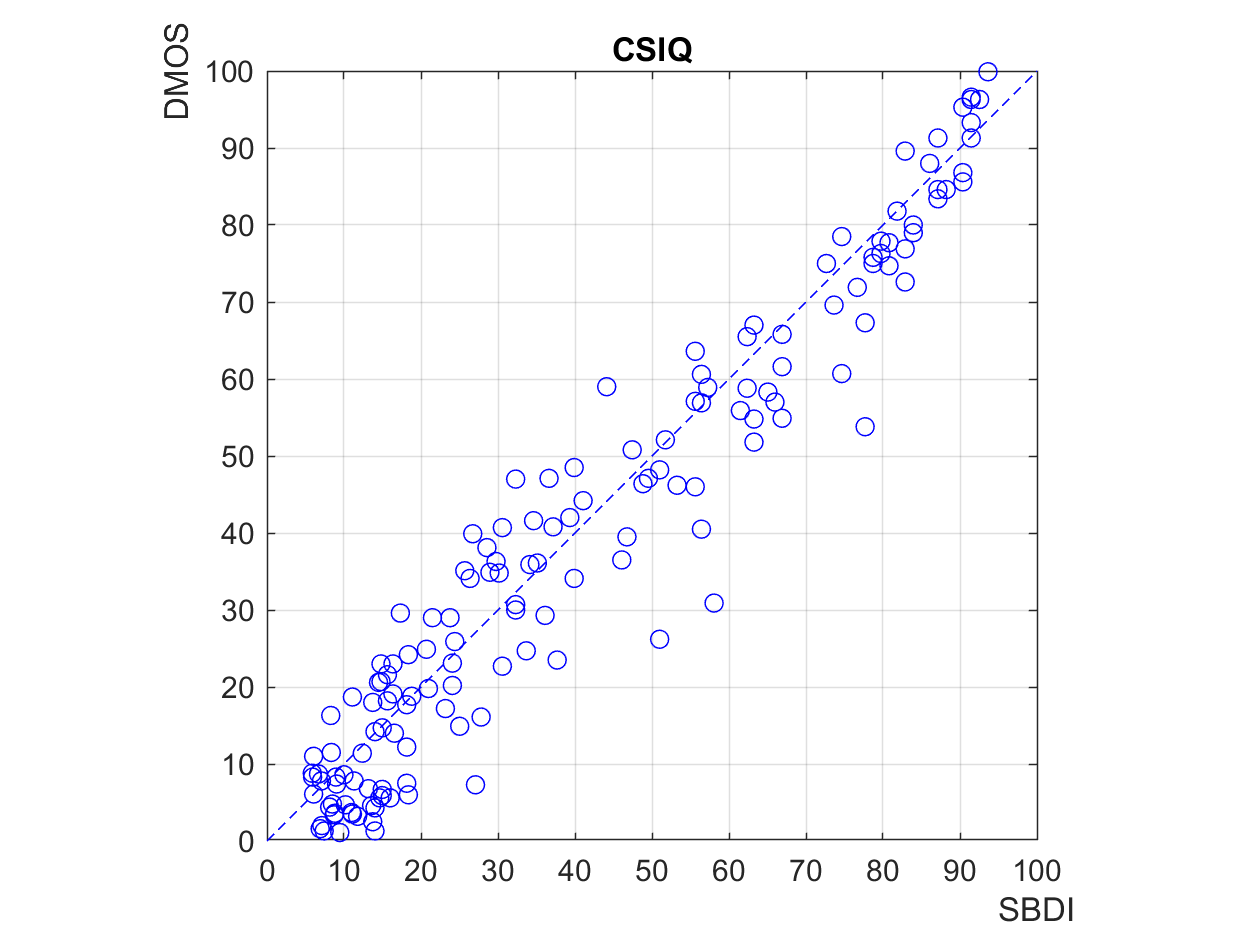}
\includegraphics[width=3.1in]{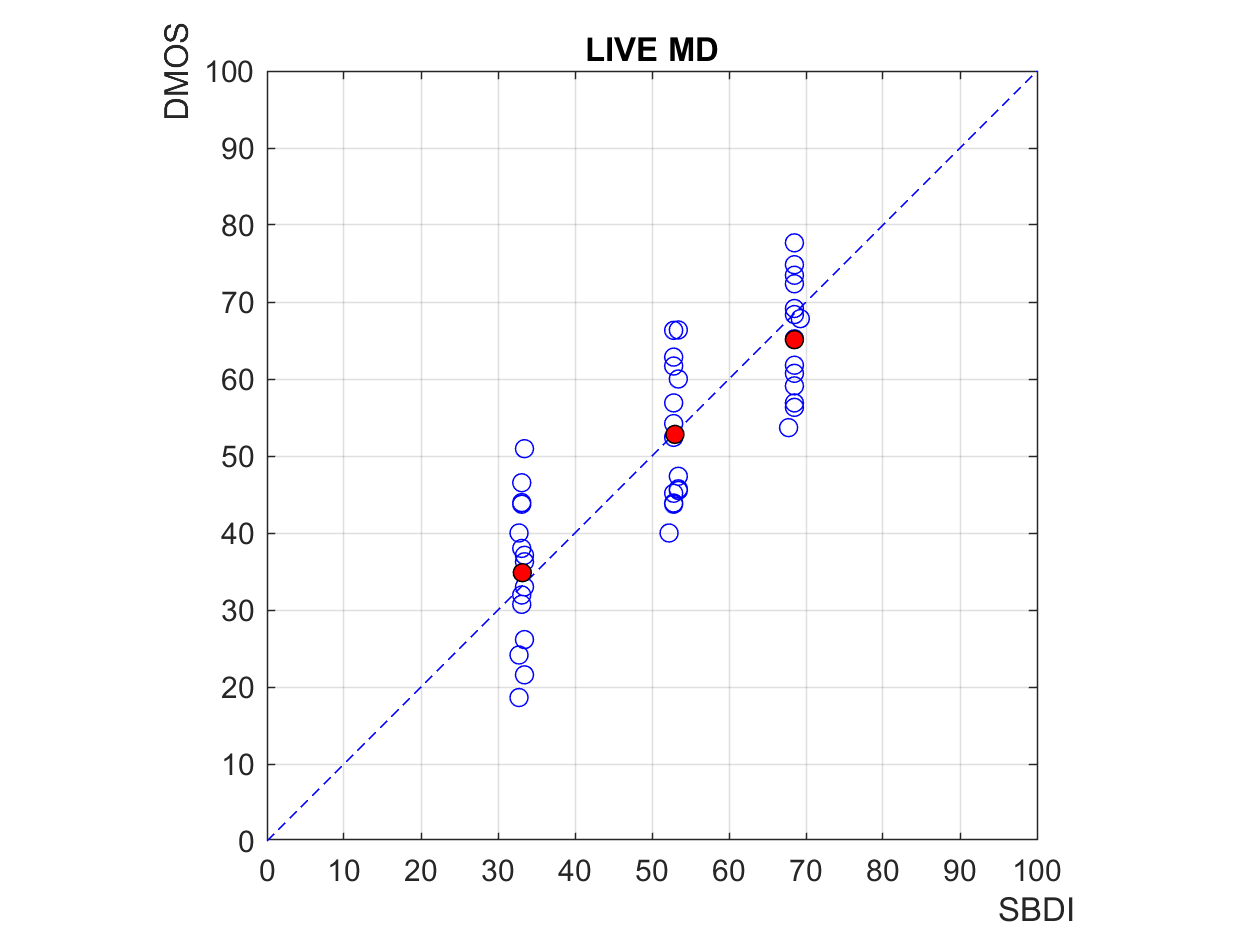}
\caption{The scatterplots of the DMOS ratings versus the corresponding SBDI values predicted by the model for each blurred image in the different databases. Averages with respect to images are indicated by filled circles in the TID2013 and LIVE MD scatterplots.}
\label{fig8}
\end{figure*}

To this purpose, the following parametric “Scaled Blur Discomfort Index” (SBDI) is defined from \eqref{eqn:epsilon1}:
\begin{equation}
SBDI\doteq a\left[1-\sqrt{\frac{1}{1+\left(\gamma^2\displaystyle\frac{s_B}{s_G}\right)^2}}\right]
\label{eqn:SBDI}
\end{equation}
where the gain $a$ fixes the \emph{scoring scale} \citep{SHEIKH06}, and $\gamma$ is the \emph{viewing distance} defined as
\begin{equation}
\gamma\doteq\frac{\delta_0}{\delta}
\label{eqn:gammaratio}
\end{equation}
where $\delta_0$ is \emph{nominal viewing distance} i.e. the distance from which the density of pixels projected on the retina matches the previously assumed density of the receptors (60/degree) and $\delta$ is the viewing distance adopted in the experiment.

Differently stated, $\gamma$ equals the ratio between the number of pixels viewed within one degree at distance $\delta_0$ (60 pixels) and the number of pixels viewed within one degree at distance $\delta$.

The role of the distance parameter is understood considering that the VRF spread on the image projected on the retina is $\displaystyle\frac{s_G}{\gamma}$, i.e. it increases proportionally with the viewing distance. Conversely, the projection of the spread of the blur applied to the observed image is $\gamma$$s_B$, i.e., it is inversely proportional to the viewing distance.

Unless specified, the parameters $a$ and $\gamma$ can be determined from data by regression.

\subsection{The “subjective quality loss” experiments}
\label{sec:The “subjective quality loss” experiments}

The essential features of the employed IQA databases are illustrated below.

The LIVE Image Quality Assessment Database Release 2 (DBR2) \citep{SHEIKH06B} reports the quality ratings of 779 distorted versions of 29 reference images (included 145 blurred images) from about 23 subjects. Ratings of subjective quality loss with respect to reference images were expressed on a DMOS (Difference of Mean Opinion Score) scale ranging from 0 (perfect quality) to 100 (bad quality) using a \emph{double stimulus} strategy.

The Tampere Image Database 2013 (TID2013) \citep{PONOMARENKO13} contains 3000 distorted images, including 125 blurred images. Quality ratings were collected in five independent labs and on the internet using more than 300 subjects. They were asked to select the \emph{best image between two distorted images} in direct comparison to the reference image. The average quality scores were expressed on a Mean Opinion Score (MOS) scale ranging from 0 (bad quality) to 9 (perfect quality).

The Computational and Subjective Image Quality Database (CSIQ) \citep{LARSON10} contains 30 reference images and 866 distorted versions, including 150 blurred images. The database includes 5000 ratings of 25 subjects, and the ratings, obtained by \emph{comparative ratings between different images}, are reported in DMOS units.

The LIVE Multiply Distorted Image Quality Database (LIVE MD) \citep{JAYARAMAN12} contains 15 reference images and 405 distorted images, including 45 blurred images, whose quality was rated by 37 subjects. The study was conducted using a \emph{single stimulus with hidden reference} strategy, using DMOS scores.

All databases contain \emph{natural images}, i.e., images representing \emph{natural scenes}, except the image i25 of the TID2013 database (which was used as a benchmark in Fig.\ref{fig3}).

In the graphs of Fig.\ref{fig7} the subjective quality loss prediction curves for the blurred images of the different databases are superposed to the empirical DMOS values, plotted versus the normalized spread ${s_B}/{s_G}$. These empirical DMOS data represent \emph{average scores} of the pool of the observers. In addition, in the TID2013 and LIVE MD cases, averages with respect to the sample images are also indicated by filled circles.

%\begin{figure*}[!ht]
%\centering
%\includegraphics[width=3.1in]{blursigma_DBR2.png}
%\includegraphics[width=3.1in]{blursigma_TID2013_mean.png}
%\includegraphics[width=3.1in]{blursigma_CSIQ.png}
%\includegraphics[width=3.1in]{blursigma_LIVEMD_mean.png}
%\caption{The empirical DMOS values of the subjective quality loss for the blurred images of the databases versus the normalized blur spread compared to the values predicted by the theoretical model (dashed curves). Averages with respect to the sample image contents are indicated by the filled circles in the TID2013 and LIVE MD scatterplots.}
%\label{fig7}
%\end{figure*}

In the TID2013 database, ratings are available as MOS values. The DMOS values were inferred considering that the best MOS ratings do not exceed 7.5 (see Fig.20 of \citep{PONOMARENKO13}). Therefore, posing $MOS=7.5$ in correspondence to $DMOS=0$ and $MOS=0$ in correspondence to $DMOS=100$ yields
\begin{equation}
DMOS=(100/7.5)\cdot (7.5-MOS)\; .
\label{eqn:DMOS_TID2013}
\end{equation}

In the CISQ database, the DMOS was normalized between its minimum and maximum empirical value.

The blur values of the LIVE MD database were not available. They were estimated through a regularized spectral division of the blurred images with the unblurred ones.

The fitting of the empirical data with the theoretical prediction of the subjective quality loss is substantially linear, as evidenced by the scatterplots of the DMOS empirical ratings versus the predicted ones for all the images contained in the different databases (Fig.\ref{fig8}). In the TID2013 and LIVE MD scatterplots, averages with respect to the sample images are indicated by filled circles.

%\begin{figure*}[!ht]
%\centering
%\includegraphics[width=3.1in]{blurDMOSideal-ref_DBR2.png}
%\includegraphics[width=3.1in]{blurDMOSideal-ref_mean_TID2013.png}
%\includegraphics[width=3.1in]{blurDMOSideal-ref_CSIQ.png}
%\includegraphics[width=3.1in]{blurDMOSideal-ref_mean_LIVEMD.png}
%\caption{The scatterplots of the DMOS ratings versus the corresponding SBDI values predicted by the model for each blurred image in the different databases. Averages with respect to images are indicated by filled circles in the TID2013 and LIVE MD scatterplots.}
%\label{fig8}
%\end{figure*}

In Table \ref{table:parameters} the most relevant data about the experimental validation of the model are resumed. The Pearson Linear Correlation Coefficient (PLCC) and the Root Mean Square Error (RMSE) between theoretical and empirical data are also provided. The “claimed” normalized viewing distances were calculated with the information about the experimental settings provided by the respective authors\footnote{The viewing distances of CISQ and LIVE MD are slightly (10\%) underestimated. The maximum absolute difference is within ten centimeters. However, the “physical” viewing distance does also depend on the physical dimension of the pixels of the screen.}. The sign “$-$” stands for “not available”\footnote{The Authors are grateful to Prof. Ponomarenko for providing details about the blur settings in the TID2013 database in a personal communication.}.

\begin{table}[htb]
\caption{Summary of the experimental verification for the four databases and $s_G=2.5$}
\centering
\renewcommand{\arraystretch}{1.}
\setlength\tabcolsep{2pt}
\begin{tabular}{@{}|c|c|c|c|c|c|}
\hline\noalign{\smallskip}
DATABASE & $a$ & $\frac{\delta}{\delta_0}$ & $\frac{\delta}{\delta_0}$ & PLCC & RMSE\\
 &  estimated & claimed & estimated &  & \\
\noalign{\smallskip}
\hline
\noalign{\smallskip}
LIVE DBR2 & $93$ & $0.46/0.57$ & $0.57$ & $0.96$ & $5.44$\\
TID2013 & $80$ & $-$ & $0.43$ & $0.92$ & $6.82$\\
CSIQ & $98$ & $0.65$ & $0.60$ & $0.97$ & $ 7.47$\\
LIVE MD & $107$ & $0.84$ & $0.76$ & $0.83$ & $8.57$\\
\hline
ALL & $-$ & $-$ & $-$ & $0.95$ & $6.44$\\
\noalign{\smallskip}
\hline
\noalign{\smallskip}
\end{tabular}
\label{table:parameters}
\end{table}

\subsection{The “blur discomfort” experiments}
\label{sec:The “blur discomfort” experiments}

In \citep{OHARE13} the results of some experiments aimed to investigate the relationship between visual discomfort judgments and image manipulations causing blur are reported. In particular “Experiment 3” regards \emph{natural images}.

In these experiments sixty natural images were taken from a database whose images pertain to two general categories: distant natural scenes and closeups natural scenes. In particular, ten images from the first category whose spectral energy has a mean radial frequency decay of the kind $\displaystyle\left(\frac{1}{\rho^\beta}\right)^2$ with $\beta=1.39$, and ten images from the second category, characterized by $\beta=0.95$, were selected. The scope of this diversity was to see if visual discomfort judgments would depend on deviation from ideal spectral decay of natural scenes ($\beta=1$).

During experimental sessions, thirteen subjects were asked to formulate \emph{discomfort judgments} following a \emph{pairwise comparison} strategy.

\begin{figure}[!ht]
\centering
\includegraphics[width=3.1in]{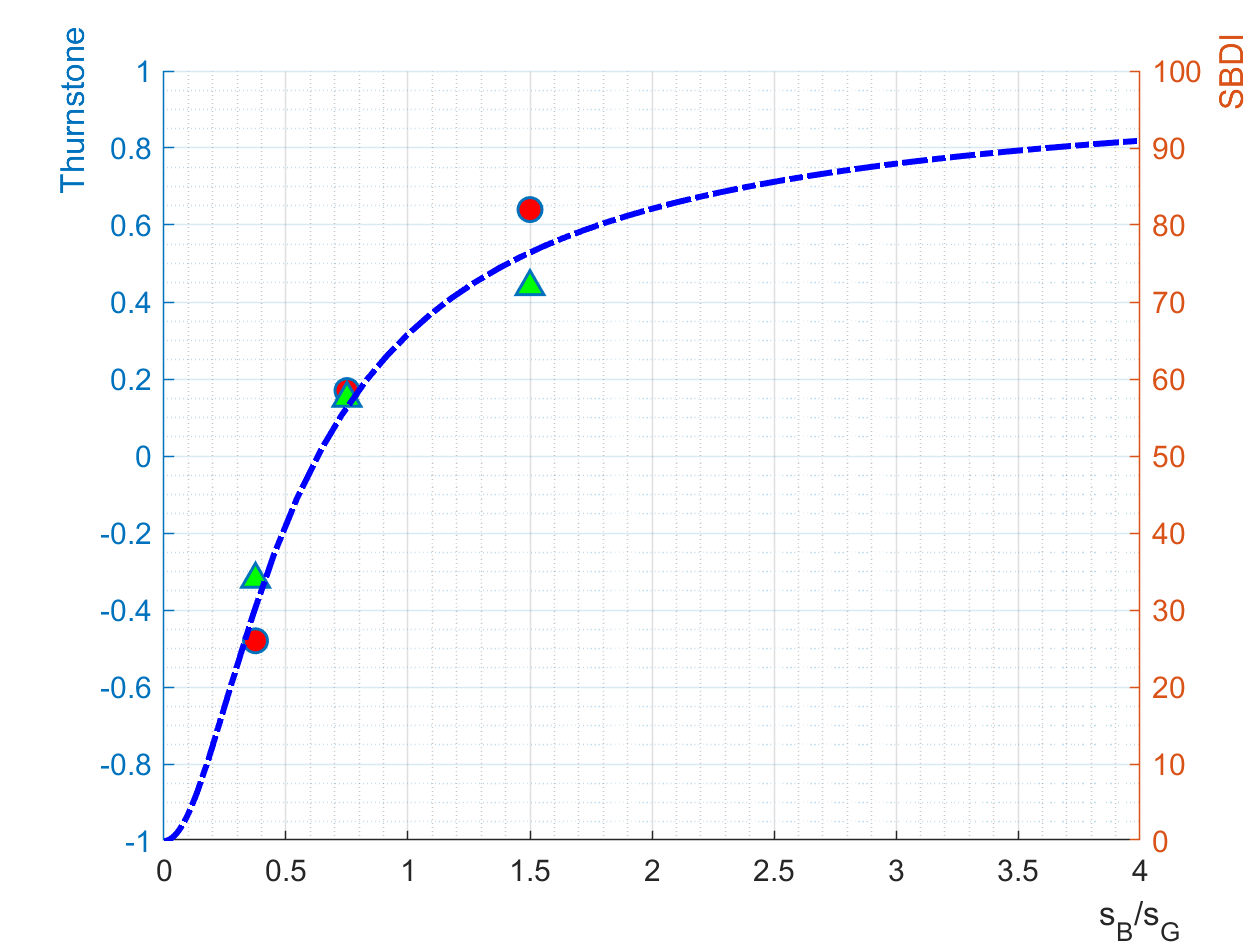}
\caption{The scatterplots of the subjective discomfort ratings versus the normalized blur spread for the images in \citep{OHARE13}. The triangles refer to the distant natural scene category, and the circles to the closeups scene category.}
\label{fig9}
\end{figure}

Differently from the preceding experiments, ratings are averaged not only over the pool of observers, but also over groups of images. One average regards the distant natural scenes, and the other one regards the closeups natural scenes. As in the TID2013 and in the LIVE MD experiments, few (three) Gaussian blur values were employed, with standard deviations 8, 16, and 32 \emph{cycles/degree} in the spatial frequency domain, corresponding respectively to $s_B=3.75$, $s_B=1.875$ and $s_B=0.9375$ \emph{arcmin} in the spatial domain. In the Fig.\ref{fig9} the results of these experiments are reported, along with the discomfort predicted by the SBDI index. The results were expressed in the Thurnstone scale \citep{TSUKIDA11}, according to the method employed for calculating the discomfort scores, and then converted into the discomfort scale of the SBDI by the following affine transformation:
\begin{equation}
SBDI=50\cdot (Thurnstone+1)\; .
\label{eqn:Thurnstone}
\end{equation}
The viewing distance, determined by regression, is $\frac{\delta}{\delta_0}=0.6$, corresponding to $\gamma=1.66$.

The scatterplots of the discomfort ratings versus the normalized Gaussian blur spread are displayed in Fig.\ref{fig9} for the groups of images contained in the database, along with the superimposed theoretical prediction curve. The triangles refer to the distant natural scene category, and the circles to the closeups scene category.

\begin{figure}[!ht]
\centering
\includegraphics[width=3.1in]{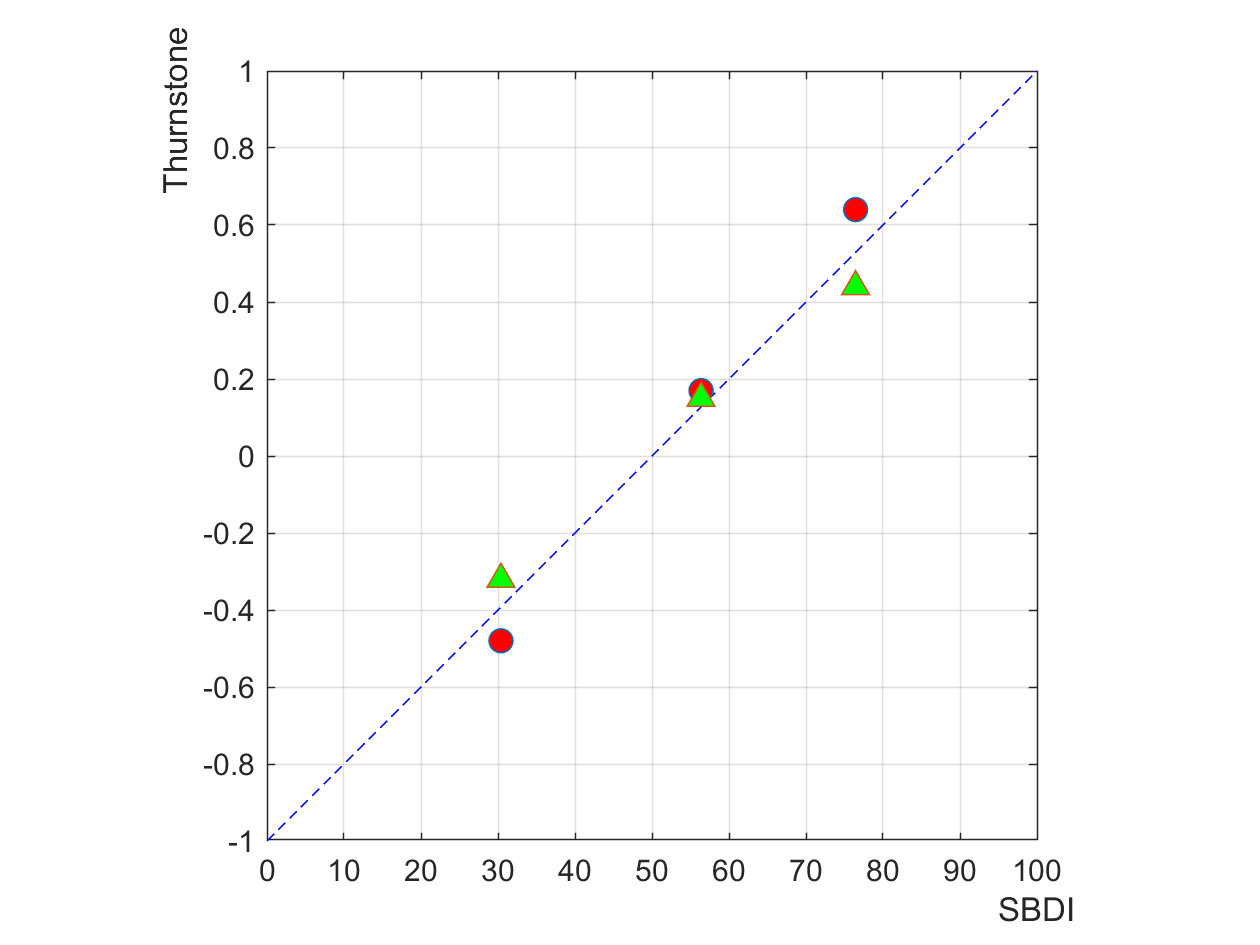}
\caption{The subjective discomfort ratings versus their theoretical predictions for the images in \citep{OHARE13}. The triangles refer to the distant natural scene category, and the circles to the closeups scene category.}
\label{fig10}
\end{figure}

The fitting of the empirical data with the theoretical prediction of the subjective quality loss is substantially linear, as evidenced by the scatterplots of the empirical ratings versus the predicted ones for all the images contained in the database (Fig.\ref{fig10}).

The Pearson Linear Correlation Coefficient (PLCC) between the average ratings and the predicted SBDI is 0.98. These results support the argued strict relationship between the subjective quality loss ratings and the visual discomfort ratings caused by blur.

\section{Remarks}
\label{sec:Remarks}

From the IQA experiments, the judgments expressed by pool of observers for the different natural images appear moderately scattered around the SBDI curve, as indicated by the Root Mean Square Error values of Tab.\ref{table:parameters}. This fact is relevant, considering the large variety of the images contained in the databases. This relatively small dispersion can be explained by the adaptive nature of the image exploration, which tends to focus on the details of prominent interest \citep{BADDELEY06}.

This appears even more evident looking at the results of the “blur discomfort” experiments. Therein, one of the purposes was to verify if the discomfort judgments depend on the deviations from ideal statistics of natural images. The average ratings for selected categories of images characterized by markedly different spectral decay are so close each other (see Fig.\ref{fig10}), that in \citep{OHARE13} the authors concluded that the difference among these spectral decays is not important to blur discomfort judgments.

This evidence underlines the validity of the attribution of the ideal spectrum of natural images to the subset of the visited details. This assumption, made in Sect.4, supports the cognitive approach followed herein.

%\begin{equation}
%\left|D_N(\rho,\vartheta)\right|^2\Rightarrow F\frac{1}{\rho^2}.
%\label{eqn:GBD}
%\end{equation}
%where the factor F is defined in \eqref{eqn:Fcoeff}.

However, the image content still influences the blur discomfort, as shown by the said moderate, but not negligible, scattering of the subjective ratings in the IQA databases, and as suggested by the common experience. The identification of essential quantitative attributes of images producing different blur discomfort in the presence of the same amount of optical blur is a matter of current investigation.

\section{Conclusion}
\label{sec:Conclusion}

The adoption of a polar separable complex-valued receptive field model, and of a visual information loss criterion, led to compact theoretical formulas for the prediction of the blur visual discomfort for natural scenes, exhibiting a good predictive power faced to several independent experimental data.

\begin{itemize}
\item{From a general scientific viewpoint, the paper proposes that the blur discomfort reflects the dissatisfaction of basic \emph{cognitive needs} of fine localization of the objects in the visual scene.}
\item{From a system theoretical viewpoint, the approach provides a \emph{functional model} of the discomfort phenomenon due to the optical blur, aimed to quantitatively predict psychophysical findings irrespective of the underlying mechanisms.}
\item{From a technical viewpoint, the results presented here may provide \emph{coarse estimates} of the discomfort caused by undesired blur, for optical correction in natural vision, and for calibration of image reproduction apparatus.}
\end{itemize}

Finally, these results could help in developing IQA methods.


%% \section*{References}

%%       \citet{<label>} ==> Jones et al. [21]
%%       \citep{<label>} ==> [21]
%%

\bibliographystyle{elsarticle-harv}
\bibliography{imageprocessing}


%% \begin{thebibliography}{00}


%% \bibitem[ ()]{}

%% \end{thebibliography}
\end{document}